\documentclass{article}

\usepackage[utf8]{inputenc}

\usepackage{amsmath, amssymb}
\usepackage{color}
\usepackage{graphicx}
\usepackage{amsmath}
\usepackage{amsthm}
\usepackage{amssymb}
\usepackage{amsfonts}
\usepackage{color}
\usepackage{xcolor}
\usepackage{graphicx}
\usepackage{caption}
\usepackage{mathrsfs}
\usepackage[utf8]{inputenc}
\usepackage[T1]{fontenc}
\usepackage{cancel}
\usepackage{comment}
\usepackage{tikz}
\usepackage{floatrow}
\usepackage{mathtools}
\DeclarePairedDelimiter\ceil{\lceil}{\rceil}

\usepackage{stackrel}

\definecolor{FG}{rgb}{0.0, 0.5, 0.0}
	
\usepackage{amsmath,amsthm,amssymb,setspace,enumitem,epsfig,titlesec,verbatim,color,array,eurosym,multirow,blkarray,lscape,bm,nicefrac,comment,soul,graphicx,MnSymbol,wasysym,hyperref,subfigure,multirow}
\usepackage[top=2.5cm,left=2.8cm,right=2.8cm,bottom=3.4cm]{geometry}

\usepackage{ulem}
\usepackage{lineno}

\usetikzlibrary{decorations.pathreplacing,angles,quotes}

\newtheorem{theorem}{Theorem}
\numberwithin{theorem}{section}

\newtheorem{lemma}[theorem]{Lemma}
\newtheorem{proposition}[theorem]{Proposition}

\definecolor{G1}{rgb}{0.0, 0.5, 0.0}

\newcommand{\tz}{\tilde{z}_+}

\allowdisplaybreaks

\title{An alternative delayed population growth difference equation model}
\author{Sabrina H. Streipert\footnote{$^1$ streipes@mcmaster.ca} \,and Gail S. K. Wolkowicz\footnote{wolkowic@math.mcmaster.ca}}

\date{\today}

\begin{document}

\maketitle

\begin{abstract}
 We propose an alternative delayed population growth difference equation model based on a modification of the Beverton--Holt recurrence,  assuming a delay  only in  the growth contribution that takes into account that those individuals that die during the delay, do not contribute to growth.  The model introduced differs from existing delay difference equations in population dynamics, such as the  delayed logistic difference equation, which  was formulated as a discretization of the Hutchinson model.  The analysis of our delayed difference equation model identifies an  important  critical delay threshold. If the time delay exceeds this threshold, the model predicts that the population will go extinct for all non-negative initial conditions
and if it is below this threshold, the population survives and  its size converges to a positive  globally asymptotically stable equilibrium that is decreasing  in size as the delay increases.
 Firstly, we obtain the local stability results by exploiting the special structure of powers of the Jacobian matrix. Secondly, we show that local stability implies global stability using two different techniques. For one set of parameter values, a contraction mapping result is applied, while for the remaining set of parameter values, we show that the result follows by first proving that the recurrence structure is  eventually monotonic in each of its arguments.
\end{abstract}

{\it keywords:
Logistic growth; Beverton--Holt;  Pielou model; Difference equations; Global stability;
Extinction threshold;  Componentwise monotonicity; Spectral radius of Matrix power;}


\section{Introduction}


The logistic growth model is a well studied differential equation, introduced by Verhulst \cite{Verhulst1838} in the context of modelling population growth. A discretization of the Verhulst model can be obtained by applying the Euler method to the logistic differential equation and is often referred to as the logistic difference equation, see for example \cite{May_2001, Smith_1968}.
Robert May \cite{May_1974} popularized this discrete version of the Verhulst model, also known as the logistic map, which contributed significantly to the mathematical study of chaos. 
This model was however criticized biologically as solutions can become negative and given its potential for chaotic behavior, not possible in the continuous logistic model, hence referring to it as the ``discrete counterpart'' does not seem appropriate. To overcome the possible negativity of solutions, a recurrence derived under the assumption that the fraction of surviving individuals is given by an exponential function is  considered to be the appropriate discretization  by some authors \cite{macfadyen1963, May_1974}. 
In this work however, we modify yet another discretization of the logistic model in order to include the effect of delay on growth, namely the Beverton--Holt model,  also known as the Pielou equation.  We refer to  this 
new model as a delayed logistic difference equation, since the Beverton--Holt model was originally derived in \cite{BH} under the assumption of an underlying logistic growth model, and authors such as \cite{BohnerWarth, brauer2001, Pielou1, Pielou2} argue that the Beverton--Holt model is a discretization of the logistic  differential equation, since it preserves  most of its properties. \\


Despite its simplicity, the Beverton--Holt equation is used in resource management to model populations, especially in fisheries science \cite{freeman2014,  haddon2011modelling, hilborn1992quantitative, SHARMA2005}. Naturally, simple mathematical models often inherit implicit assumptions on processes, for example both may assume uniform spatial movement. As these assumptions are not necessarily satisfied  in real-world systems, model predictions should be interpreted carefully, dependent on the level of violation of these assumptions. There are however benefits in applying simple models. One reason is that these models are usually  more tractable and are often well studied. Simple population models, such as the Beverton--Holt model, are preferred for use in data-limited species assessment models \cite{Dichmont2016, Froese2016,  Punt2015,  Rosenberg2017, JABBA, Worm2009}. Furthermore, more complex models are frequently constructed using  such simple models as building blocks. For example, age-structure population models often use the Beverton--Holt model as the recruitment function \cite{haddon2011modelling}.\\

To improve a model, one may start to refine assumptions, one by one, to capture more realistic features. For example, an implicit assumption of simple population models is a rather uniform behavior of the population, meaning that all individuals are assumed to behave alike. This is rarely the case since for example, one  expects differences in  traits based on sex and age. To model these trait variations, a natural extension of such models is therefore the addition of variables or by including age structure. Such models are usually higher-dimensional mathematical models. In the age-structured case, one needs to follow the age-distribution of the population throughout time. While age-structured population models may make more precise predictions, they do require the collection of specific age dependent data that is not always economically or biologically feasible \cite{hilborn1992quantitative}. In \cite{Deriso1980}, Deriso suggested incorporating delay in models as a  compromise between  simple and the age-structured population models, a technique that was extended by others in, for example,  \cite{Doonan1987, Schnute1985, Schnute1987}. While the simplicity of the model structure is preserved in these corresponding delay models, the contribution of different age classes to the change in biomass can be considered without keeping track of the precise age-distribution. In this work, we use this technique to implement a time until positive fecundity, which is a crucial age-structured property. More precisely, we derive a population model based on the Beverton--Holt model under the assumption that it takes $\tau$ time units to reach fertile age and consequently promote population growth.\\


The above arguments led to the inclusion  of delay in continuous and discrete population models. A popular  modification of the Verhulst model is the delay logistic differential equation (Hutchinson or Wright model). The Hutchinson model has been extensively studied by several authors, see for example \cite{Cushing1977, Gopalsamy1992,Jaquette2017, MacDonald1978, Nisbet1982, VANDENBERG20187412}, despite certain questionable properties. More precisely, the size of the equilibrium of the Hutchinson model is independent of the delay and is  globally asymptotically stable if and only if the product of the growth rate and the time delay is bounded by the rather un-intuitive bound of $\frac{\pi}{2}$, see \cite{VANDENBERG20187412}. For parameter values that do not satisfy this bound, solutions of the Hutchinson model exhibit another unreasonable property that nontrivial periodic solutions  persist independent of the length of the delay. 
While the Hutchinson model was derived assuming a delay in the per-capita growth rate, the alternative delay differential equation formulated in \cite{ArinoWangWolkowicz2006} includes a delay solely in the growth process 
 and takes into consideration the fact that those individuals that die during the delay, do not contribute to growth. The authors in \cite{ArinoWangWolkowicz2006} show that their model predicts that the population dies out if the delay exceeds a certain threshold and  converges to a globally asymptotically stable equilibrium with size that decreases as the delay increases. This behavior seems more reasonable for populations in  natural ecosystems. The recurrence introduced in this work is derived using the same assumptions as in \cite{ArinoWangWolkowicz2006} and exhibits similar properties. It can therefore be considered as the discrete analogue of \cite{ArinoWangWolkowicz2006}.\\


The  discrete delay population model that we propose also differs from the popularized delay logistic difference equation introduced in \cite{Pielou1} and discussed by many authors, including \cite{CAMOUZIS2007,Kocic2010, Ladas1993, Arora2004, Merino2007, Pielou1, Pielou2}.  As in our model, the discrete delay model in \cite{Pielou1} also exploits the relation between the continuous logistic model and the Beverton--Holt model, but was introduced as a discretization of the Hutchinson equation instead of the alternative formulation in \cite{ArinoWangWolkowicz2006}. The model in \cite{Pielou1} can be criticized for the same reasons as Hutchinson's model, because it exhibits the same questionable behavior described above, see \cite{Liz_2019, Ladas1993, Ladas_2001, Ladas1992}. 
In contrast, solutions of the model that we  propose converge to a positive equilibrium with size depending on the length of the delay for small delay and  converge to zero (i.e., the population goes extinct) if the delay exceeds a critical threshold. \\

The paper is organized as follows. In Section 2, we derive the discrete delay model by modifying the classical Beverton--Holt model. In Section 3, we begin our analysis of the proposed model with the local stability of the trivial and the unique positive equilibria by exploiting the structure of the Jacobian matrix and its powers. In that process, we identify a critical threshold for the delay. We continue studying the global dynamical behavior. We prove that for positive initial conditions, if the delay exceeds the critical threshold, then the trivial equilibrium is globally asymptotically stable. Instead, if the delay falls below the threshold, then the population survives and converges to a positive equilibrium that decreases in size as the delay increases. Thus, the dynamics of our discrete model mirrors most of the qualitative behavior predicted by the continuous model in \cite{ArinoWangWolkowicz2006}. 

 The main difference between the predictions of these two models is that
in the continuous model, if the initial data is either entirely above or entirely below the positive equilibrium, solutions converge to it  monotonically.  We provide numerical simulations to show that this is not the case for the discrete model. Another simulation illustrates how the dynamics differ for certain choices of parameters, and hence show why a different technique was needed to prove the global stability of the positive equilibrium. Finally, in the conclusion in Section 4, we summarize our results and  highlight the differences between the dynamical behavior of our modified Beverton--Holt model and two related models: its continuous analogue introduced in \cite{ArinoWangWolkowicz2006} and its underlying submodel, the Beverton--Holt model.


\section{Derivation of a discrete delay growth model}

In this section, we derive a delayed logistic model by identifying the growth and decline contributions in the Beverton--Holt model before incorporating a time lag in the growth component  taking into consideration that those that die during the delay do not contribute to growth.  A similar technique was applied in the derivation of the logistic delay differential equation introduced in \cite{ArinoWangWolkowicz2006}.
The classical Beverton--Holt model is given by
\begin{equation}\label{BHclassic}
y_{t+1} = \frac{\rho K y_t}{K + (\rho-1)y_t},
\end{equation}
with $K \in \mathbb{R}^+$, representing the carrying capacity, $\rho >1$, the proliferation rate, and $y_t$,  the population at time $t$. Recurrence \eqref{BHclassic} was obtained in \cite{BH} by solving the logistic growth model and relating the solution evaluated at time $t+T$ to the solution at time $t$. The parameter $\rho$ was introduced by substituting $\rho=e^{rT}$ for $r>0$, resulting in $\rho>1$, as outlined in \cite{BH}. The recurrence \eqref{BHclassic} can be normalized using the variable transformation $z_t = \frac{y_t}{K}$ resulting in 
\begin{equation}\label{eq9}
z_{t+1} = \frac{z_t}{\frac{1}{\rho} + \frac{(\rho-1)}{\rho}z_t} = p_t z_t,
\end{equation}
where 
\begin{equation}\label{eq11star}
p_t := \frac{1}{\frac{1}{\rho} + \frac{\rho-1}{\rho}z_t},
\end{equation}
can be interpreted as the survival probability. Following the reasoning in \cite{ArinoWangWolkowicz2006}, we assume that the survival probability depends on growth, death, and intraspecific competition. Then, \eqref{eq11star} reveals that the term $\frac{\rho-1}{\rho}$ determines the decline due to intraspecific competition and $\frac{1}{\rho}<1$ is the sum of the death and growth contribution. The growth contribution  can  generally  be expressed as $\frac{1}{\rho}=1+b-a$, where $a>0$ is the growth component and $b>0$ the death component. Since $\rho>1$, this implies that $a-b \in (0,1)$, i.e., the value of the actual growth component exceeds the value of the death component. 

To highlight each of the three components: growth, death, and intraspecific competition, we therefore, express the survival probability  \eqref{eq11star} as
\begin{equation}\label{eq10star}
p_t=\frac{1}{1-(a-b) + c z_t}.
\end{equation}
Expression \eqref{eq10star} is decreasing in $b$ and $c$, due to death and competition,  and  is increasing in $a$, representing the growth contribution. We note that the distinction of intraspecific competition, growth, and death follows the approach in \cite{ArinoWangWolkowicz2006}, where the authors consider these three components  before implementing a delay solely in the growth component of the rate of change. In this work, we proceed similarly and consider a delay only in the growth contribution.

The simple species model \eqref{BHclassic} describes the relation between non-overlapping generations. That is, individuals of the ``old'' generation reproduce at time $t$ to form the ``new'' generation. After one time unit, the ``old'' generation is replaced by the ``new'' generation and the cycle repeats. The time unit can therefore be understood as the length of the reproductive cycle, which is equal to  the generation time. In \cite{Deriso1980}, Deriso justified the use of delay models, among other reasons, to describe the dynamics of species where the reproductive cycle is not equal to the generation time. This is the case, for example, when newborn individuals do not contribute to reproduction immediately, but rather reach fecundity after $\tau$ reproductive cycles. Then, the group of fecund individuals at time $t+1$ not only depends on the fecund individuals at time $t$, but also on individuals that reach fecundity for the first time at time $t+1$.

Based on the survival probability in \eqref{eq10star}, the individuals exposed to death and competition follow the recursion 
\[w_{t+1} = \frac{1}{1+b+cw_t}w_t.\]
This can be solved in the same way as for the Beverton--Holt model, i.e., by multiplying both sides by the denominator to obtain
$$w_{t+1} + bw_{t+1} + cw_t w_{t+1} = w_t,$$
and hence
$$\Delta w_t = w_{t+1} - w_t = -w_{t+1}(b+cw_t).$$
Substituting $v_t = \frac{1}{w_t}$ (for $w_t \neq 0$) to obtain
$$\Delta v_t = \frac{-\Delta w_t}{w_t w_{t+1}} =  bv_t +c,$$
yields a first order linear difference equation with the solution given in \cite{kelley2001difference} by
\begin{eqnarray*}
v_t &=& (1+b)^{t-t_0}v_{t_0} +\sum_{i=t_0}^{t-1} (1+b)^{t-i-1}c,\\
&=& (1+b)^{t-t_0}v_{t_0} +\frac{c(1+b)^t}{1+b} \sum_{i=t_0}^{t-1} \left( \frac{1}{1+b}\right)^{i}.
\end{eqnarray*}
Using the formula for the sum of a geometric series,
\begin{eqnarray*}
    v_t &=& (1+b)^{t-t_0}v_{t_0} +\frac{c(1+b)^t}{1+b} \left\{ \frac{1-\frac{1}{(1+b)^{t}}}{1-\frac{1}{1+b}}  - \frac{1-\frac{1}{(1+b)^{t_0}}}{1-\frac{1}{1+b}}\right\},\\
 &=& (1+b)^{t-t_0}\left(v_{t_0} +\frac{c}{b}\right)-\frac{c}{b}.
\end{eqnarray*}
Returning to $w_t$,  yields
$$w_t =  \frac{bw_{t_0}}{(1+b)^{t-t_0}\left(b + cw_{t_0} \right)-cw_{t_0}}.$$
Setting $t_0=t-\tau$, yields the fraction of individuals at time $t-\tau$ that survive to time $t$ as
\begin{equation}\label{DefF}
w_t =  \frac{bw_{t-\tau}}{(1+b)^{\tau}\left(b + cw_{t-\tau} \right)-cw_{t-\tau}}:=F(w_{t-\tau}).
\end{equation}

The surviving fraction is now used in the recurrence 
$$z_{t+1} = p_t z_t = \frac{1}{1+b+cz_t - a}z_t,$$
where $b+cz_t$ determines the decay and $a$ the growth. Rearranging, we obtain 
\begin{equation}\label{newrec}
z_{t+1}(1+b+cz_t) - az_{t+1} =z_t.
\end{equation}
Recalling that $a$ is the growth rate, we identify $az_{t+1}$ as the growth contribution of the Beverton--Holt recurrence. 
If, for example, fecundity is reached after $\tau>1$ reproductive cycles, it is reasonable to consider a delay in the growth contribution. We therefore assume that the growth contribution is  proportional to the 
(fecund) population at time $t-\tau$ that survive until time $t+1$.
Thus, we replace $az_{t+1}$ in \eqref{newrec} by $aF(z_{t+1-\tau})$, where $F$ determines the fraction of $z_{t-\tau+1}$ that survives $\tau$ units, given in \eqref{DefF}, to obtain
$$z_{t+1}(1+b+cz_t) - aF(z_{t+1-\tau}) =z_t.$$
Solving this for $z_{t+1}$ yields the delay difference recurrence
$$z_{t+1} =
\frac{1}{1+b+cz_t}\left( z_t + aF(z_{t-\tau+1}) \right).$$
The fecund individuals at time $t+1$, denoted by $z_{t+1}$, is therefore given by the sum of the surviving  fecund individuals $z_t$ and the surviving individuals reaching fecundity for the first time, expressed by $aF(z_{t-\tau+1})$. The surviving probability $1/m(z_t)$, then multiplies the sum $z_t+aF(z_{t-\tau+1})$.

By \eqref{DefF}, the recurrence we obtain is
\begin{align}\label{eq10}
z_{t+1} =\frac{1}{m(z_t)}\left\{ z_t +   \frac{ab \, z_{t-\tau+1}}{M(z_{t-\tau+1})}\right\}=:  H(z_t,z_{t-\tau+1}),
\end{align}
with 
\begin{equation}\label{Mm}
m(x) := 1+b+cx, \quad \quad 
  M(x) := b\beta  + (\beta-1)cx,  \, \quad \mbox{ and } \quad \beta := (1+b)^{\tau}.
\end{equation}

If $\tau=0$, no time lag exists and the generation cycle is equal to the reproductive cycle. Then, \eqref{eq10} reduces to 
$$z_{t+1} = \frac{z_t}{1+b+cz_t} + \frac{a}{1+b+cz_t}z_{t+1},$$
which, after rearranging terms, yields
$$z_{t+1} = \frac{z_t}{1+b-a+cz_t}.$$
This is the equation used to derive the model, which is by \eqref{eq9}--\eqref{eq10star} an equivalent expression for the Beverton--Holt model when $1+b-a=\frac{1}{\rho}$ and $c=\frac{\rho-1}{\rho}$. This recurrence is well established and has been extensively studied, see for example \cite{BH,BohnerWarth,Ladas1993,haddon2011modelling, Pielou1, Pielou2}.  
Therefore,  throughout this paper, we assume $\tau>0$. Although the derivation of \eqref{eq10}  assumed certain  relationships between the parameters $a,b,c$, the recurrence remains valid for arbitrary parameter choices of $a,b,c\geq 0$. 
For this reason  we consider the dynamics of  \eqref{eq10} with \eqref{Mm} in the  following sections, requiring only that $a,b,c\geq 0$. This is consistent with the study of the Beverton--Holt model. Even though the derivation by Beverton and Holt in 1957 lead to specific domains for the model parameters \cite{BH}, the recurrence remains valid for arbitrary positive parameters. Thus, the Beverton--Holt model under the assumption of arbitrary positive parameter values, also known as the Pielou equation, became the focus of many studies.

\section{Dynamics of the discrete delay  difference equation}

In this section, we present results concerning the dynamics of the  discrete delay recurrence equation \eqref{eq10}
for $t, \tau\in \mathbb{N}=\{1,2,3,\ldots, \}, t\geq \tau>0$, and initial conditions 
\begin{equation*}
\vec{z}_0=(z_0, z_1, \ldots, z_{\tau-1}) \quad \mbox{with} \quad  z_i\geq 0 \quad \mbox{for}\quad i=0, 1, \ldots, \tau-1.
\end{equation*}
We consider \eqref{eq10} with \eqref{Mm} and  $a,b,c >0$, unless explicitly stated otherwise. We justify the focus on $a,b,c>0$ by noting that if $a=0$ or $b=0$ or $c=0$, no positive equilibrium exists. 

We start our analysis with some basic results about the existence of fixed points, before continuing to the discussion of their local and global stability. The proofs are given in the appendix.

We define a \emph{critical delay} $\tau_c \in \mathbb{R}$ as
\begin{equation}\label{Tau}
\tau_c:=\frac{\log(\frac{a}{b})}{\log(1+b)} \end{equation}
and remind the reader that for recurrences $\tau \in \mathbb{N}=\{1,2,\ldots, \}$. Therefore, the inequality $\tau\geq \tau_c$ is understood as $\tau \in [\tau_c, \infty) \cap \mathbb{N}_0$, which we express henceforth by $\tau \geq \ceil{\tau_c} = \min\{n\geq \tau_c \, : \, n \in \mathbb{Z}\}$ for $\tau_c \in \mathbb{R}$. Similarly, the inequality $\tau<\tau_c$ is expressed as $\tau<\ceil{\tau_c}$.
We point out that for $\tau \in \mathbb{N}$
\begin{equation}\label{arel}
    \tau\geq \ceil{\tau_c} \quad  \iff \quad \tau \geq \tau_c \quad   \iff \quad a\leq  b\beta 
\end{equation}
for $\beta$ defined in \eqref{Mm}. The equivalence \eqref{arel} implies that for $\tau \in \mathbb{N}$, the inequality $\tau<\ceil{\tau_c}$ is equivalent to $\tau< \tau_c$ which is equivalent to $a>b \beta$. This relation is extensively exploited in proofs of this section's claims.

We also point out that if $\tau_c \in (0,1]$, then $\tau < \ceil{\tau_c}$ is only satisfied for $\tau=0$ in which case \eqref{eq10} reduces to the classical Beverton--Holt model.

The first result addresses the positivity of solutions as formulated in the proceeding lemma.
\begin{lemma}\label{Lpos}
Let $z_t$ be a solution of \eqref{eq10}.
If $z_i=0, \ i=0,1,\dots,\tau-1$, then $z_i=0$ for all $i\in\mathbb{N}_0=\{0,1,2,\ldots\}.$
If $z_s>0$ for at least one  $s \in \{0, 1, \ldots, \tau-1\}$,  then $z_t> 0$ for all $t\geq \tau+s$.
\end{lemma}

\begin{theorem}\label{TmEx}
If $\tau \geq \ceil{\tau_c}$, then $\tilde{z}_0\equiv 0$ is the only non-negative equilibrium. If $\tau < \ceil{\tau_c}$, then there also exists   a unique positive equilibrium 
\begin{equation}\label{FP}
\tilde{z}_{+} = \frac{- b(2\beta -1) + \sqrt{\left( b(2\beta -1)\right)^2 +4b(a-b\beta) (\beta -1)}}{2c (\beta -1)}.
\end{equation}
\end{theorem}

Equation \eqref{FP} reveals the importance of $a>b\beta$, which is by \eqref{arel} equivalent to $\tau<\tau_c$, to assure the positivity of $\tz$. A positive equilibrium can therefore only exist if the delay is below a critical upper bound, else the population is doomed to go extinct. 
Since $\beta$ depends on $\tau$, the positive equilibrium $\tz$ is a function of the delay. In fact, $\tz$ is monotone decreasing in the delay which ultimately yields an upper bound, as formulated below.

\begin{lemma}\label{upbnd}
Let $\tau<\ceil{\tau_c}$. Then $\tilde{z}_+$ given in \eqref{FP} is decreasing in $\tau$ and
$\tilde{z}_{+}\leq \frac{a-b}{c}$. 
\end{lemma}

As in the continuous delay logistic model in \cite{ArinoWangWolkowicz2006}, the positive unique equilibrium is decreasing in size as the delay increases. This dependency of the equilibrium on the delay also highlights a difference to the existing discrete delay Beverton--Holt model in \cite{Pielou1}, in which the equilibrium is independent of the delay.

To study the local stability, we rewrite \eqref{eq10} as $\vec{w}_{t+1} = G(\vec{w}_t)$, where $\vec{w}_t  \in \mathbb{R}^{\tau}$ with the $i^{th}$ component $w_{t,i}=(\vec{w}_{t})_i =z_{t-i+1}$ for $i=1, 2, \ldots, \tau$. 
Linearization yields
$$\vec{w}_{t+1} \approx J \vec{w}_t$$
where $J \in \mathbb{R}^{\tau \times \tau}$ is the Jacobian given by  
\begin{equation}\label{Jac}
J = \begin{bmatrix} \frac{1+b}{m^2(w_{t,1})}-\frac{ac}{m^2(w_{t,1})}\frac{bw_{t,\tau}}{M(w_{t,\tau})} & 0 & 0 & \ldots & 0 & \frac{ab}{m(w_{t,1})}\frac{b\beta}{M^2(w_{t,\tau})}  \\
1 & 0 & 0 & \ldots & 0 & 0\\
0 &1  & 0 & \ldots & 0 & 0\\
\vdots & \vdots & \vdots & \vdots & \vdots & \vdots\\
0 &0  & 0 & \ldots & 1 & 0\\
\end{bmatrix}.
\end{equation}
The Jacobian \eqref{Jac} is of the special form 
$$A=\begin{bmatrix} x & 0 & 0 & \ldots & 0 & y  \\
1 & 0 & 0 & \ldots & 0 & 0\\
0 &1  & 0 & \ldots & 0 & 0\\
\vdots & \vdots & \vdots & \vdots & \vdots & \vdots\\
0 &0  & 0 & \ldots & 1 & 0\\
\end{bmatrix}=\begin{bmatrix} x\vec{e}_1 & y\\
I_{\tau-1} & 0 \end{bmatrix}$$
where $\vec{e}_1 = (1, 0, \ldots, 0)$ is the standard basis vector  in  $\mathbb{R}^{\tau-1}$, $x,y \in \mathbb{R}$, and  $I_{\tau-1}$ is the  identity matrix in $\mathbb{R}^{\tau-1 \times \tau-1}$. This generic matrix form has, for general $x,y \in \mathbb{R}$, pleasant properties that can be exploited.

\begin{lemma}\label{MaPower1}
Let $x,y \in \mathbb{R}$, $N \in \{2,3,\ldots\}$. Consider any  matrix of the form
\begin{equation}\label{Ma}
A = \begin{bmatrix} x\vec{e}_1 & y\\
I_{N-1} & 0 \end{bmatrix}\in \mathbb{R}^{N\times N}.
\end{equation}
where $\vec{e}_1 = (1, 0, \ldots, 0)$ is the standard basis vector  in  $\mathbb{R}^{N-1}$  and  $I_{N-1}$ is the  identity matrix in $\mathbb{R}^{N-1 \times N-1}$.
Then 
\begin{equation}\label{EqMP}
    A^{N} = \begin{bmatrix} 
x^{ N}+y & xy & x^2y & x^3y & x^4y & \ldots & x^{N-1}y\\
x^{N-1}&y & xy & x^2y & x^3y &  \ldots & x^{N-2}y\\
x^{N-2}&0&y & xy & x^2y &   \ldots & x^{N-3}y\\
\vdots & \vdots & \vdots & \vdots & \vdots & \vdots & \vdots \\
x&0&0 & 0 & 0 &   \ldots & y\\
\end{bmatrix}.
\end{equation}
\end{lemma}

This special structure of $A^N$ serves the purpose of identifying the row-sum norm of the last row as the corresponding matrix norm as stated in the next lemma.

\begin{lemma}\label{MaPower}
Consider $A \in \mathbb{R}^{N \times N}$ of the form  \eqref{Ma},  with $|x|+|y|<1$. Then 
$$\sum_{i=1}^N |A^{N}_{k,i}|\leq \sum_{i=1}^N |A^{N}_{j,i}|, \quad \mbox{for} \quad 1\leq k <j\leq N, $$
and
$$\|A^N\| = \max \limits_{1\leq j \leq N}\sum_{i=1}^N |A^{N}_{j,i}| = \sum_{i=1}^N |A^{N}_{N,i}|=|x|+|y|<1,$$
where $A^N_{j,i}$ denotes the  entry in the $j$th row and $i$th column of $A^N$.
\end{lemma}

Since the Jacobian of \eqref{eq10} is of the form \eqref{Ma}, lemmas \ref{MaPower1} and \ref{MaPower} are utilized to prove the following statements concerning the local stability of the non-negative equilibria.

\begin{theorem}\label{Tmlocal}
Consider \eqref{eq10}.
\begin{enumerate}
    \item[a)] The trivial equilibrium, $\tilde{z}_0$, is
    \begin{enumerate}
    \item[i)]  locally asymptotically stable if  $\tau \geq \ceil{\tau_c}$, and is 
     \item[ii)]  unstable if $\tau < \ceil{\tau_c}$.
     \end{enumerate}
    \item[b)] Whenever   the positive equilibrium, $\tilde{z}_+$ exists, i.e., $\tau < \ceil{\tau_c}$, 
it is locally asymptotically stable.
\end{enumerate}
\end{theorem}

Now we bring our attention to the global stability of \eqref{eq10} and being  with the global asymptotic stability of the trivial equilibrium. 

\begin{theorem}\label{TmGlob0}
If $\tau \geq \ceil{\tau_c}$, then $\tilde{z}_0$ is globally asymptotically stable for solutions with non-negative initial conditions.
\end{theorem}

Hence, if the positive equilibrium does not exist, that is, if the delay exceeds the critical delay $\tau_c$, then the population goes extinct over time. This is reasonable remembering that the delay determines the growth contribution and therefore, if the time span to reach fecundity is longer than the critical time $\tau_c$, the decline component dominates which leads to the species' extinction. We point out that this is consistent with the corresponding continuous model in \cite{ArinoWangWolkowicz2006}, where the trivial equilibrium is globally stable if the delay exceeds a critical delay.

It remains to discuss the case when $\tau<\tau_c$ and the unique positive equilibrium exists and point out that the global stability of the trivial solution was obtained using a contraction argument. For $\tau<\tau_c$, we distinguish between $c\tz>1$ and $c\tz\leq 1$ and use for each case a different technique to discuss global stability. The case of $c\tz\leq 1$ is in fact related to the derivation of the model \eqref{eq10} because the recurrence was derived in Section 2 for $a-b=\frac{1}{\rho}$ for $\rho>1$, which implies $a-b \in (0,1)$. By Lemma \ref{upbnd}, $\tilde{z}_+\leq \frac{a-b}{c}$, which results, for $a-b \in (0,1)$, in $\tz\leq \frac{1}{c}$. In that case when $c\tz\leq 1$, solutions have specific properties. As formally stated below, solutions with initial conditions that are all below $\tz$ remain below $\tz$ and solutions with initial conditions all above $\tz$ remain above $\tz$. This property also holds for the classical Beverton--Holt model \eqref{BHclassic} for $\rho>1$ (i.e., $\tau=0$).
On the other hand, if some of the $\tau$ initial conditions are above and some are below $\tz$, then solutions of \eqref{eq10} can oscillate about the equilibrium $\tz$ but are bounded by the minimum and maximum value of the initial conditions. We will show that these properties ultimately lead to the global asymptotic stability of $\tz$ if $\tau< \tau_c$.

 We emphasize that the case of $c\tz\leq 1$ corresponds in general to the following parameter relation.
 
\begin{proposition}\label{alowbnd}
\begin{equation}\label{acause}
  c\tz\leq 1  \quad \iff \quad   a \leq   \frac{1+b}{b}\left( b\beta+\beta-1\right)   
\end{equation}
\end{proposition}

\begin{lemma}\label{mono}
Let $\tau<\ceil{\tau_c}$ and $c\tilde{z}_+\leq 1$. If  $z_i\leq \tilde{z}_+$  for $i=0,1,\ldots, \tau-1$, then $z_t\leq \tilde{z}_+$ for all $t \geq 0$. If $z_i\geq \tilde{z}_+$ for $i=0,1,\ldots, \tau-1$, then $z_t\geq \tilde{z}_+$ for all $t\geq 0$. 
\end{lemma}

\begin{lemma}\label{Bound}
Let $\tau<\ceil{\tau_c}$ and $c\tilde{z}_+\leq 1$. If $z_i>0$ for $i=0,1,\ldots, \tau-1$, then 
\begin{equation}\label{eq:bound}\min\{ z_0,z_1,\ldots, z_{\tau-1},\tilde{z}_+\}\leq z_t\leq \max\{z_0,z_1,\ldots, z_{\tau-1}, \tilde{z}_+\}, \quad  \quad \mbox{for all} \quad t\geq 0.
\end{equation}
\end{lemma}

\begin{theorem}\label{TmGlobpos}
Let $\tau < \ceil{\tau_c}$. If $c\tilde{z}_+\leq 1$, then $\tilde{z}_{+}$ is globally asymptotically stable for solutions with initial conditions $\vec{z}_0\not \equiv \vec{0}$.
\end{theorem}

\begin{figure}[h!]
    \centering
    \includegraphics[scale=0.5]{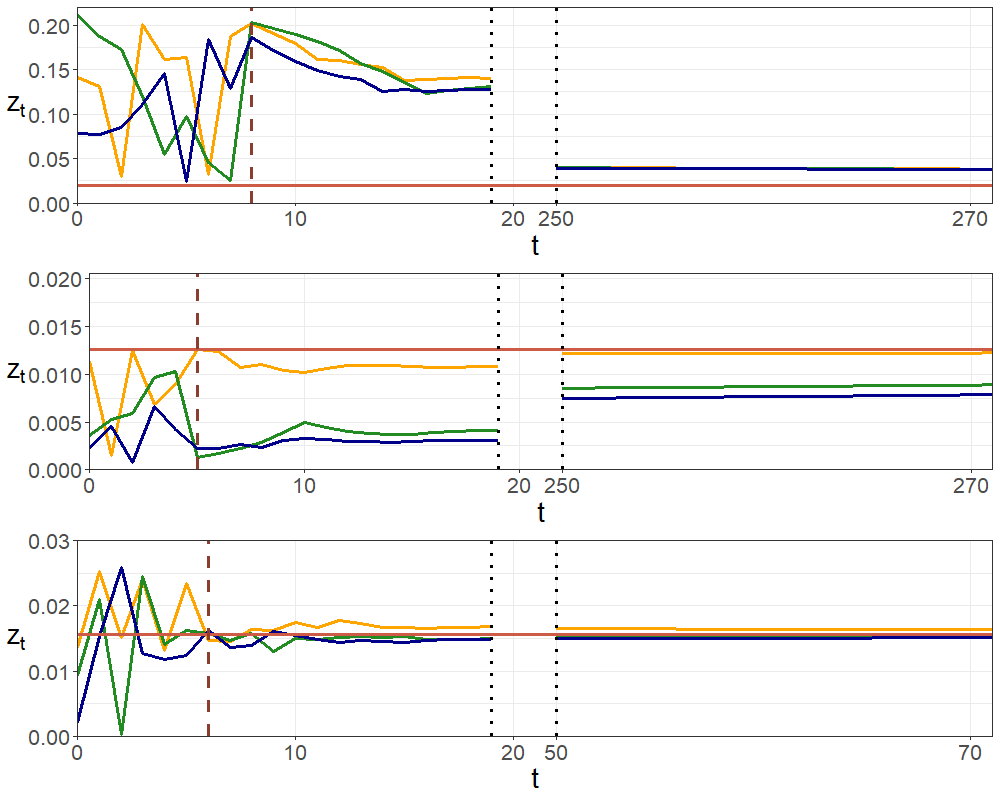}
    \caption{Behavior of three solutions for different parameters with $c\tz\leq 1$. Top panel: 
    $a=0.272, b=0.105, c=0.142, \tau=9$. 
    Middle panel: $a=0.508,  b=0.177, c =0.663, \tau= 6$. 
    Bottom panel: $a=0.884, b= 0.209, c= 0.905, \tau= 7$. 
    The values on the left of the vertical dashed line are the initial conditions. The solid horizontal line is $\tilde{z}_+$. If the initial conditions are on either side of $\tz$, the solutions seem to converge to $\tz$ faster, see bottom panel compared to the top and middle panel. }
    \label{FigD}
\end{figure}

Figure \ref{FigD} illustrates the dynamic behavior of solutions of \eqref{eq10} for three different initial conditions in the case when $c\tilde{z}_+\leq 1$. If $\tau \in \mathbb{N}$ is chosen to be less than $\tau_c$, then solutions with at least one positive initial condition converge to the positive equilibrium $\tilde{z}_+$, as per Theorem \ref{TmGlobpos}. This coincides with the global behavior for the continuous model in \cite{ArinoWangWolkowicz2006}. However, unlike the corresponding continuous model, solutions to \eqref{eq10} can be non-monotone, independent on whether all initial conditions are above $\tz$ (top panel in Figure \ref{FigD}), below $\tz$ (middle panel  in Figure \ref{FigD}), or on either side (bottom panel in Figure \ref{FigD}). Note that all of the figures in this paper were produced using the software package R \cite{R:software}. The non-monotone behavior of solutions differs from the behavior of those of the classical Beverton--Holt model for $\rho>1$, where solutions monotonically increase (decrease) to the positive equilibrium $K$ for initial conditions below (above) $K$.

\begin{figure}[h!]
    \centering
    \includegraphics[scale=0.4]{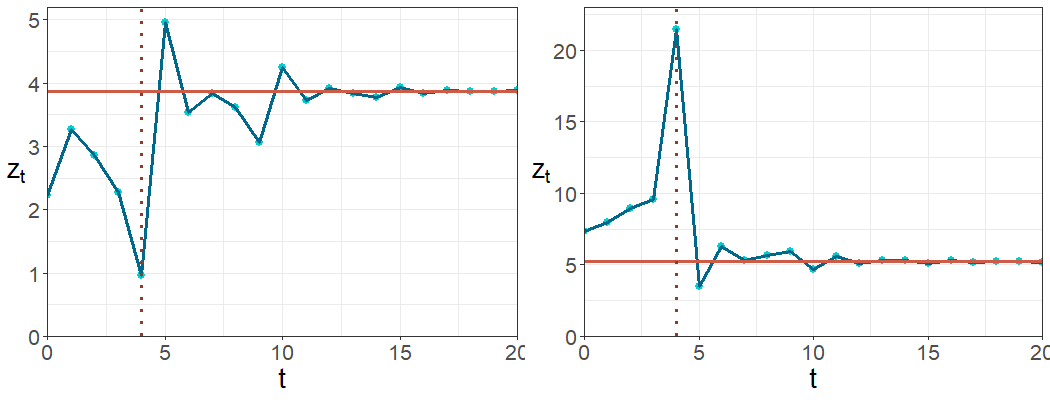}
\caption{Solutions for parameter values when $c\tz>1$, hence violating the assumption in Lemma~\ref{mono}.  The solid line represents the positive equilibrium. The dotted line identifies the last initial condition. 
{\it Left panel:} $a=38.72$, $b=0.227$, $c=0.498$, $\tau=5$, $\vec{z}_0=2.226,	3.274,	2.861,	2.269,	0.956$.
{\it Right panel:} $a=137.78$,  $b=0.640$, $c=0.417$, $\tau=5$, $\vec{z}_0= 7.368,	7.985,	8.934,	9.572,	21.444$.
Although the initial conditions are all below $\tz$ (left panel), subsequent iterates can be above $\tz$. Similarly on the right, although all initial conditions are above $\tz$, subsequent iterates can fall below $\tz$. }
    \label{Viol1}
\end{figure}

Figure \ref{Viol1} demonstrates that Lemma \ref{mono} can not be extended to the case when $c\tz>1$. Solutions with all initial conditions below $\tz$ can exceed $\tz$ eventually (left panel). Similarly, solutions  with  initial conditions that are all above $\tz$ can obtain values below $\tz$ (right panel).

Theorem~\ref{TmGlobpos} exploits the contraction mapping theorem but this technique fails if $c\tz>1$. Instead, if $c\tz>1$, the global asymptotic stability of $\tz$ if $\tau<\tau_c$ can be obtained using Theorem 1.15 in \cite{Ladas_2004}, stated in the appendix for completion. For the application of this theorem, we require some preliminary work.

\begin{proposition}\label{Lowerbnd}
Let $\tau<\tau_c$. If $c\tz>1$, then there exists $T \in \mathbb{N}$ such that for $t\geq T$,
\begin{equation}\label{Defchi}
z_t\geq \chi := \frac{b(1+b)\beta}{c(ab-(\beta-1)(1+b))} \quad \end{equation}
and $H(z_t,z_{t-\tau+1})$ is decreasing in the first variable.
\end{proposition}

\begin{proposition}\label{invariant}
Consider  $H$ defined in \eqref{eq10} and $\chi$  defined in \eqref{Defchi}. There exists $U^*>\tz$, such that for all $U\geq U^*$, 
$H:[\chi,U]\times [\chi,U] \to [\chi,U]$. 
\end{proposition}

Propositions \ref{invariant} and \ref{Lowerbnd} are fundamental in the proof of the global asymptotic stability of the positive equilibrium. 

\begin{theorem}\label{Thm_glob_pos2}
Let $\tau<\tau_c$. If $c\tz>1$, then $\tz$ is globally asymptotically stable for initial conditions $\vec{z}_0\not \equiv \vec{0}$.
\end{theorem}

Theorems \ref{Thm_glob_pos2}, \ref{TmGlobpos}, and \ref{TmGlob0} provide combined the global asymptotic stability of the nonnegative equilibria. Consequently, the positive equilibrium $\tz$  is globally asymptotically stable whenever it exists, else the trivial solution is globally asymptotically stable. 

\begin{figure}[h!]
    \centering
    \includegraphics[scale=0.35]{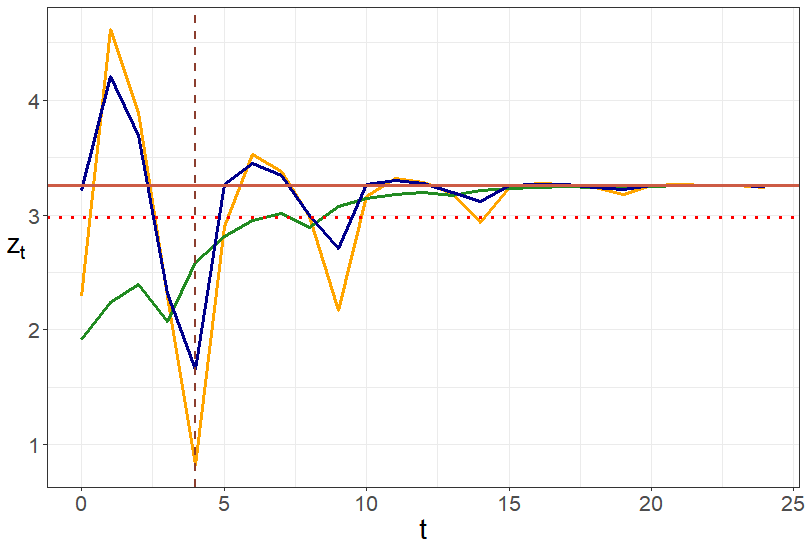}
    \caption{Behavior of three solutions to parameter combinations where $c\tz>1$ for $a=38.118, b= 0.557, c=0.313, \tau= 5,  \tz = 3.258$. The vertical dashed line separates the initial conditions from the iterations. The solid horizontal line is $\tz$. The dashed horizontal line is at the y-value $\chi=2.9819$. If $z_t>\chi$, then the function $H$ in \eqref{eq10} is increasing in both variables $z_t$ and $z_{t-\tau+1}$, else it decreases in the first variable. The figure illustrates a crucial result used in the proof of the global stability of $\tz$, namely that solutions are eventually above $\chi$.}
    \label{fig:my_label}
\end{figure}

\section{Conclusion}

In this paper, we introduced an alternative delayed Beverton--Holt model that can be viewed as the discretization of the delayed logistic model in \cite{ArinoWangWolkowicz2006}. Starting from the (classical) Beverton--Holt model \eqref{BHclassic}, the survival probability was assumed to depend on three components: growth, death, and intraspecific competition. To account for a time delay in the growth, created for example by a time lag in reaching fecundity, the recurrence was rearranged to identify the growth term. The model takes into consideration the fact that those individuals that die during the delay, do not contribute to growth. This method is consistent with the approach in \cite{ArinoWangWolkowicz2006}, where an alternative delayed logistic  differential equations model was formulated. Even though in the derivation of the delay recurrence model, we made  certain restrictions  on the  parameter values,  we studied the recurrence for arbitrary positive parameters, since the recurrence model remains mathematically valid. Since the model reduces to the classical Beverton--Holt model in the case of no delay, we focused on the model analysis when the delay $\tau>0$, that is $\tau \in \{1,2,3,\ldots, \}$.

We began the analysis of our delayed Beverton--Holt model by exploiting the special structure of the Jacobian matrix and its powers that allowed us to identify a critical threshold for the delay. We showed that the trivial solution of our model is globally asymptotically stable if the delay is bigger than this critical threshold. For the parameter values assumed in the derivation of our model, we proved the global asymptotic stability of the survival equilibrium if the delay is below the threshold using a contraction mapping argument.  We used a different technique to prove the global asymptotic stability of the positive equilibrium in the case of arbitrary positive parameter values that relies on  componentwise monotonicity.

Some of the properties of the delay Beverton--Holt model \eqref{eq10} that we introduced, are similar to those of the classical Beverton--Holt model. More specifically, for parameter values consistent with the derivation of our model in Section 2, solutions with initial conditions above (below) the unique positive equilibrium remain above (below) the equilibrium. In contrast, solutions of our model do not always converge monotonically even for parameter values consistent with the derivation of the model. This non-monotone behavior of  solutions of our model was illustrated with simulations. However, the corresponding figures also seem to indicate an eventual monotonic convergence to the positive equilibrium.

We justified that our recurrence \eqref{eq10} is an appropriate discretization of the delay logistic model introduced in \cite{ArinoWangWolkowicz2006}. Both models separate the net-growth rate into three components: growth, death, and intraspecific competition and consider a time lag only in the growth contribution and take into consideration that those members of the population that die during the delay period do not contribute to growth.  Further, both models exhibit similar dynamics dependent on a critical threshold. If the delay is below the critical threshold, solutions of both models converge to a positive equilibrium that decreases in size for increasing delay values, and converge to the trivial solution, otherwise.
We think that these model properties are reasonable for natural ecosystems and differ from the properties of the other discrete and continuous delay logistic models already mentioned in this paper. There are however, slight variations in the dynamic behavior between the solutions of the continuous and the discrete models.  In contrast to the continuous counterpart, the solutions of the discrete model do not always converge monotonically, even if the initial distribution is entirely above or entirely below the positive equilibrium, but rather can display damped oscillations about the positive equilibrium, as demonstrated in Figures \ref{FigD}--\ref{fig:my_label}.

\section*{Appendix: Proofs}\label{App}

\begin{proof}[\textbf{Proof of Lemma \ref{Lpos}}]
Clearly if all the components of the initial condition $\vec{z}_0$ equal zero, then $z_i=0$ for all $i\in\mathbb{N}_0.$   
If on the other hand,  there exists at least one $s\in\{0,1,\dots,\tau-1\}$  such that $z_s>0$, then 
since the right hand side of \eqref{eq10} is always non-negative,  
$z_{s+\tau}=\frac{z_{s+\tau-1}}{m(z_{s+\tau-1})}+a\frac{bz_s}{m(z_{s+\tau-1})M(z_{s})}>0$. 
Then $z_{s+\tau+1}\geq \frac{z_{s+\tau}}{m(z_{s+\tau})} >0$.  Similarly, $z_{s+\tau+i}>0$ for $i \geq 1$.
\end{proof}

\begin{proof}[\textbf{Proof of Theorem \ref{TmEx}}]
By \eqref{eq10} with \eqref{Mm}, a positive equilibrium $\tilde{z}_+$ satisfies the equation
$$\tilde{z}_+ =  \frac{\tilde{z}_+}{1+b+c\tilde{z}_+} + \frac{a}{1+b+c\tilde{z}_+}\cdot 
\frac{b\tilde{z}_+}{b \beta  + (\beta-1)c\tz }.$$
Rearranging and using the fact that $\tilde{z}_+\neq0$, it follows that $\tilde{z}_+$ must satisfy the quadratic equation
$$(\beta-1)(c{\tilde{z}}_+)^2 +(b\beta+b(\beta-1)) c \tilde{z}_+ +b(b\beta-a)=0.$$
By the Routh-Hurwitz condition, since $\beta>1$ for $\tau\geq 1$, there are no roots with positive real parts unless $a>b\beta$, and in this case there are two real roots, one positive and one negative.
Solving the quadratic equation, the positive real root, $\tilde{z}_{+}$,
is given by \eqref{FP}.
\end{proof}

\begin{proof}[\textbf{Proof of Lemma \ref{upbnd}}]
By \eqref{arel}, $a>\beta b>b$ and by \eqref{eq10}, if $\tau<\tau_c$, then $\tilde{z}_+$ exists and satisfies 
$$\tz = \frac{\tz}{m(\tz)} +a \frac{b\tz}{m(\tz) M(\tz)}.$$
Rearranging, we obtain
$$(b+c\tz)(b \beta  + (\beta-1)c\tz) = ab.$$
We note that by \eqref{eq10}, $\tz$ depends on $\beta$ and therefore on $\tau$. Hence, by the above, $f(\tau)=\tz$ solves
\begin{equation}\label{Delta1}
(b+cf(\tau))(b \beta  + (\beta-1)cf(\tau)) = ab.
\end{equation}
Taking the difference of \eqref{Delta1} evaluated at $\tau+1$ and $\tau$, we have 
$$(b+cf(\tau+1))(b \beta (1+b)  + c(\beta(1+b)-1)f(\tau+1))-(b+cf(\tau))(b \beta  + (\beta-1)cf(\tau)) = 0,$$
i.e., with $\Delta f=f(\tau+1)-f(\tau)$,
$$b^3\beta+bc (\beta-1)\left( \Delta f\right)+\beta b^2 c f(\tau+1) +cb \beta \left( \Delta f\right) +cb^2 \beta f(\tau+1) + c^2(\beta-1) \left(\Delta f \right)\left( f(\tau+1)+f(\tau)\right)+b \beta c^2 f^2(\tau+1)=0$$
i.e., 
$$ \Delta f = 
-\frac{b^3\beta+\beta b^2 c f(\tau+1) +cb^2 \beta f(\tau+1) +b \beta c^2 f^2(\tau+1)}{bc (\beta-1)+cb \beta+ c^2(\beta-1) \left( f(\tau+1)+f(\tau)\right)}<0$$
for $f(\tau), f(\tau+1) \geq 0$.
Therefore, $f(\tau)=\tz$ is decreasing and $\tz$ evaluated at $\tau=0$ is an upper bound for $\tz$. To obtain this upper bound, we note that for $\tau=0$,  $\beta=1$ and \eqref{eq10} reads as 
$$z_{t+1} = \frac{z_t}{1+b+cz_t}+\frac{az_{t+1}}{1+b+cz_t}.$$
Rearranging terms yields
$$z_{t+1}\left[ 1 - \frac{a}{1+b+cz_t}\right] = \frac{z_t}{1+b+cz_t},$$
which is of the Beverton--Holt type, defined earlier,
$$z_{t+1} = \frac{z_t}{1+b+cz_t-a}.$$
Its  nontrivial equilibrium is $\tilde{z} = \frac{a-b}{c}$ which is the upper bound for $\tz$ evaluated at $\tau>0$.
\end{proof}

\begin{proof}[\textbf{Proof of Lemma \ref{MaPower1}}]
Note that $A=L+B$, where $L= \begin{bmatrix} \vec{0}_{1\times N-1} & 0\\
I_{N-1 \times N-1} & 0\end{bmatrix}$ is a lower shift matrix and $B = \begin{bmatrix} x\vec{e}_1  & y\\
[0]_{N-1 \times N-1} & 0 \end{bmatrix}$. Premultiplying a matrix   by a lower shift matrix   shifts the   elements of the matrix downward by one position and replaces the top row by zeros. As a consequence, the second row of $A^{j+1} = A A^j$ is replaced by the first row of $A^j$. In general, the $k^{th}$ row, for $k=2, 3, \dots, N$ of $A^{j+1}$ is the $(k-1)^{st}$ row of $A^j$. Hence, the $k^{th}$ row of $A^{N}$, denoted by $\left(A^{N}\right)_{k,\cdot}$, is equal to the first row of $A^{N-k+1}$, denoted by $\left(A^{N-k+1} \right)_{1, \cdot}$ for $k=2, \ldots, N$.
By the above shift, $(A^{j})_{N, \cdot} = (A^{j-1})_{N-1,\cdot}$, and by the structure of $A$, we have 
\begin{equation}\label{lastrow}
\left(A^{j} \right)_{N,\cdot} = (0, 0, \ldots, 0, \underbrace{1}_{N-j}, 0, 0, \ldots, 0    ).
\end{equation}
We now claim that the first row of $A^j$ for $j=1, \ldots, N-1$ is given by
\begin{equation}\label{PMI}
\left(A^{j}\right)_{1,\cdot } =  (x^{j}, 0, \ldots, 0, \underbrace{y}_{N+1-j}, xy,  \ldots, x^{j-1}y).
\end{equation}
To justify this, we proceed using induction. First note that \eqref{PMI} holds for $j=1$, since then  \eqref{PMI} implies that
$$\left(A^{1}\right)_{1,\cdot } =  (x, 0, \ldots, 0, \ldots, 0, y),  $$
which is equal to first row of $A$. Assume now that the statement is true for $k>1$. Since the last row of $A^k$ is given in \eqref{lastrow} by $\vec{e}_{N-k} = (0, 0, \ldots, 0, \underbrace{1}_{N-k}, 0, \ldots, 0)$, we have  
\begin{multline*}
\left(A^{k+1}\right)_{1,\cdot} = \left(A A^k\right)_{1, \cdot}= (x, 0, \ldots, 0, y) A^k = x (A^k)_{1,\cdot} + y(A^k)_{N,\cdot}\\ = 
 (x^{k+1}, 0, \ldots, 0, \underbrace{xy}_{N+1-k}, x^2y,  \ldots, xx^{k-1}y) + y(0,0,\ldots, 0, \underbrace{1}_{N-k}, 0, \ldots, 0) \\= 
 (x^{k+1}, 0, \ldots, 0, \underbrace{y}_{N-k}, \underbrace{xy}_{N+1-k}, x^2y,  \ldots, x^{k}y),
\end{multline*}
confirming the claim. 
The first row of $A^N$, $N>1$, is similarly given by 
\begin{multline*}
\left(A^{N}\right)_{1,\cdot} = \left(A A^{N-1}\right)_{1, \cdot}= (x, 0, \ldots, 0, y) A^{N-1} = x (A^{N-1})_{1,\cdot} + y(A^{N-1})_{N,\cdot}\\ = 
x(x^{N-1}, y, xy, x^2y,  \ldots, x^{N-2}y) + y\vec{e}_1=(x^N +y, xy, x^2 y, \ldots, x^{N-1}y).
\end{multline*}
\end{proof}

\begin{proof}[\textbf{Proof of Lemma \ref{MaPower}}]
Let $|x|+|y|<1$. Then, for $j=1,2,\ldots, N$,
$$\sum_{i=1}^{N} |A^N_{j,i}| \leq |x|^{N-j+1}+|y|\sum_{k=0}^{N-j} |x|^k, $$
where equality holds unless $j=1$ and $\mbox{sign}(x) \neq \mbox{sign}(y)$.

For $j\in\{2,\dots N\}$, taking the absolute value of all of the terms of $A^N$,   subtracting adjacent rows and noting that most of the terms cancel, and then factoring $|x|^{N-j+1}$, it follows that
$$\sum_{i=1}^{N} (|A^N_{j,i}|- |A^N_{j-1,i}|) \geq |x|^{N-j+1} (1-(|x|+|y|))>0,$$
if $|x|+|y|<1,$
where the first inequality is an equality for $j\geq 3 $.

This implies that the larger the row, the larger the row-sum, and hence the last row has the largest row sum   so that
$$\|A^{N}\| = \max \limits_{1\leq j\leq N} \sum_{i=1}^{N} |a_{j,i}| = \sum_{i=1}^{N} |a_{N,i}|=\sum_{i=1}^{N}|A^{N}_{N, i}|=|x|+|y| <1.$$
\end{proof}

\begin{proof}[\textbf{Proof of Theorem \ref{Tmlocal}}]
\it{a)} Evaluating the Jacobian of \eqref{eq10}, $J$, at $\vec{z}_0$, gives \eqref{Jac}. This is of the form \eqref{Ma} with $x=\frac{1}{1+b} \in (0,1)$ and $y=\frac{a}{\beta (1+b)}\in (0,1)$.  
    
\it{a)\, i)}  First, we consider the case $\tau>\ceil{\tau_c}\geq \tau_c$ and as pointed out in \eqref{arel}, $a<b \beta$.
By Lemma \ref{MaPower1}, $J^{\tau}$ is given by \eqref{EqMP}, and 
$$x+y = \frac{1}{1+b}+\frac{a}{\beta (1+b)}=\frac{\beta +a}{\beta (1+b)}<1. $$
 By Lemma \ref{MaPower}, with $J$ playing the role of $A$, it follows that $\|J^{\tau}\|<1$. Since the spectral radius, $\rho(J)\leq \|J^{\tau}\|^{1/\tau}$ for any consistent norm, we obtain the asymptotic stability of $\tilde{z}_0$. 

Next we  consider the case, $\tau=\tau_c$, then $a=b\beta$.  Select any $\epsilon>0$. If $z_i \in [0,\epsilon]$ for $i=0, 1, \ldots, \tau-1$. Then 
$$0\leq z_{t+1} \leq \frac{z_t}{1+b+cz_t}+\frac{b z_{t-\tau+1}}{1+b+cz_t} \leq \frac{1+b}{1+b+cz_t} \epsilon \leq \epsilon. $$
Hence,  $z_{t+1} \in [0,\epsilon]$, for all $t\geq 0$. This implies not only that $\tilde{z}_0$ is stable, but also that the sequence $z_{t}$ is bounded. 

By Lemma~\ref{Lpos},  $\liminf_{t\to \infty} z_t\geq 0$. To prove that $\tilde{z}_0$ is attractive, we proceed using proof by contradiction. Suppose that $\bar{z} = \limsup_{t\to \infty} z_t>0.$ 

Recalling $H$ in \eqref{eq10}, the partial derivatives satisfy
\begin{align}
    \frac{\partial H(u,v)}{\partial u} &=\frac{1+b}{(1+b+cu)^2} - \frac{ b^2c\beta v}{(1+b+cu)^2 [b \beta + (\beta-1)cv]} \label{eq:partialH1} \\
    &\geq \frac{1}{(1+b+cu)^2} (1+b-bc v) \nonumber \\
        \frac{\partial H(u,v)}{\partial v} &= \frac{b^3 \beta^2 }{(1+b+cu)(b \beta  + (\beta-1)cv)^2}>0. \label{eq:partialH2}
\end{align}

If $\epsilon<\frac{1+b}{bc}$, then $\frac{\partial H(u,v)}{\partial u}>0$ for $v \in [0,\epsilon]$, Therefore, $H$ is monotone increasing in both variables for $u,v \in [0,\epsilon]$ and
\begin{align*}
\bar{z}&=\limsup z_{t+1}= \limsup H(z_t, z_{t-\tau+1}) \leq H(\limsup z_t, \limsup z_{t-\tau+1})\\
&=H(\bar{z}, \bar{z}) = \frac{\bar{z}}{(1+b+c\bar{z})}+\frac{b^2  \beta \bar{z}}{(1+b+c\bar{z})(b \beta  + (\beta-1)c\bar{z})}\\[2mm]
&= \frac{b \beta (1+b) + (\beta-1)c\bar{z}}{ b \beta  (1+b)+ (\beta-1)c\bar{z}+b(\beta-1)c\bar{z} +c\bar{z}b \beta  +c^2\bar{z}^2(\beta-1) } \bar{z} < \bar{z},
\end{align*}
 contradicting  $\bar{z}>0$.  Hence,  $\tilde{z}_0$ is locally asymptotically stable. 

\it{a) \, ii)} Next, we prove that $\tilde{z}_0$ is unstable when $\tau< \ceil{\tau_c}$,  and as pointed out in \eqref{arel}, $a> b \beta$. Since the characteristic equation  of the Jacobian $J$ evaluated at $\tilde{z}_0$ is given by
$$P(\lambda) = \lambda^{\tau} - \frac{1}{1+b} \lambda^{\tau-1} - \frac{a}{\beta (1+b)}=0$$ 
and $P(1)<0$, but $\lim_{\lambda\to \infty}P(\lambda) =\infty$, there is a real root $\lambda>1$, and hence $\tilde{z}_0$ is unstable.

\it{b)}  Let $\tau<\ceil{\tau_c}$, then by \eqref{arel}, $a>b\beta$. By Lemma~\ref{TmEx}, $\tilde{z}_+$ exists and is unique.  Since $\tilde{z}_+$ is an equilibrium of   \eqref{eq10}, we obtain 
\begin{equation}\label{E_E}
ab =  (b+c\tilde{z}_+)(b \beta + (\beta-1)c\tz) = (m(\tz)-1)M(\tz).
\end{equation}
The Jacobian $J$ of \eqref{eq10}  evaluated at $\tilde{z}_+$ is of the form \eqref{EqMP} with
\begin{align*}
x &= \frac{1+b}{m^2(\tz)} -\frac{c\tilde{z}_+}{m^2(\tz)} \cdot \frac{ab}{b \beta + (\beta-1)c\tilde{z}_+}
\stackrel{\eqref{E_E}}{=} \frac{1+b}{m^2(\tz)} - \frac{c\tilde{z}_+}{m^2(\tz)}(b+c\tilde{z}_+)=\frac{1-c\tilde{z}_+}{m(\tz)},
\end{align*}
and
\begin{align*}
y& =  \frac{ab^2\beta}{m(\tz)M^2(\tz)}  \stackrel{\eqref{E_E}}{=} 
  \frac{\beta(b+c\tilde{z}_+)^2}{am(\tz)}
= \frac{\beta(m(\tz)-1)^2}{am(\tz)}.
\end{align*}
Note that $y>0$.

We show that 
\begin{equation}\label{Eqxpy}
    |x| + |y| = \left|\frac{1-c\tilde{z}_+}{m(\tz)}\right| +  \left|\frac{\beta(m(\tz)-1)^2}{am(\tz)}\right|<1,
\end{equation}
since then, by Lemma \ref{MaPower}, $\|J^{\tau}\|<1$. 

Firstly, if $c\tilde{z}_+=1$, then $x=0$ and $$|x|+|y|=y= \frac{\beta(m(\tz)-1)^2}{am(\tz)} \, {\stackrel{\eqref{E_E}}{=}} \, \frac{b\beta (m(\tz)-1)}{m(\tz)(b\beta+(\beta-1)c\tilde{z}_+)}<1.$$

Secondly, we assume that $0 <c\tilde{z}_+<1$. 
Then $x$ is also positive and the inequality in
 \eqref{Eqxpy} is equivalent to 
$$\beta \frac{(m(\tz)-1)^2}{a}<m(\tz)-1+c\tilde{z}_+.$$
Since, $m(\tz)-1=b+c\tilde{z}_+$ and using \eqref{E_E}, this inequality can be rewritten as 
$$\beta \frac{(m(\tz)-1)b}{  [b \beta  + (\beta-1)c\tilde{z}_+]}<m(\tz)-1+c\tilde{z}_+,$$
or, equivalently,
 $$\beta (m(\tz)-1) b<(m(\tz)-1)[b \beta  + (\beta-1)c\tilde{z}_+]+c\tilde{z}_+[b \beta  + (\beta-1)c\tilde{z}_+].$$
 Since this is clearly satisfied, it follows that 
when $c\tilde{z}_+<1$,  \eqref{Eqxpy} also holds.

Thirdly, if  $c\tilde{z}_+>1$, then $-x>0$ and 
$$|x|+|y|=-x+y = \frac{c\tilde{z}_+-1}{m(\tz)}+\frac{\beta (m(\tz)-1)^2}{am(\tz)} = \frac{a(c\tilde{z}_+-1)+\beta (m(\tz)-1)^2}{am(\tz)}.$$
To show \eqref{Eqxpy}, we show
$$a(c\tilde{z}_+-1)+\beta (m(\tz)-1)^2 < am(\tz) = a(1+b)+ac\tilde{z}_+.$$
Cancelling terms and recalling that $m(\tz)-1=b+c\tilde{z}_+$, we have the equivalent form
$$\beta (b+c\tilde{z}_+)^2 < a(2+b) \, \stackrel{\eqref{E_E}}{=} \,\frac{(2+b)}{b}(b+c\tilde{z}_+)(b \beta + (\beta-1)c\tz).$$
Simplifying this inequality yields
$$\left(\frac{2+b}{b}\right) c \tilde{z}_+ < \beta(b+c\tilde{z}_+)\left(\frac{2+b}{b}-1\right).$$
Further simplification results in
$$(2+b)c\tilde{z}_+   < 2 \beta (b+c\tilde{z}_+) =2 (1+b)^{\tau} (b+c\tilde{z}_+).$$
Since 
$$2 (1+b)^{\tau} (b+c\tilde{z}_+)\geq 2 (1+b) (b+c\tilde{z}_+) > (2+b)c\tilde{z}_+,$$
 \eqref{Eqxpy} holds. 

Therefore, in all three cases,  \eqref{Eqxpy} holds independent of the sign of $c\tz-1$ and so by Lemma \ref{MaPower}, $\| J^{\tau}\|<1$.
Since the spectral radius $\rho( J)\leq \| J^{\tau}\|^\frac{1}{\tau}<1$,  $\tilde{z}_+$ is locally asymptotically stable.
\end{proof}

\begin{proof}[\textbf{Proof of Theorem \ref{TmGlob0}}]
By \eqref{eq10},  for $z_t\geq 0$ for all $t\geq 0$, 
\begin{align*}
|H(u,v)| &=\Big| \frac{u}{1+b+cu} + \frac{a}{1+b+cu}\cdot \frac{bv}{b \beta  + (\beta-1)cv}\Big|\\
&\leq \Big| \frac{u}{1+b+cu} \Big|+ \Big|\frac{a}{1+b+cu}\cdot \frac{bv}{b \beta  + (\beta-1)cv}\Big|\\
&\leq 
\Big| \frac{u}{1+b}\Big| +\Big| \frac{a}{1+b}\Big| \cdot \Big| \frac{bv}{b \beta }\Big|\\
&\leq \alpha \|u\|_{\infty},
\end{align*}
with $\alpha=\Big| \frac{1}{1+b}+\frac{a}{(1+b)\beta}\Big|$ and  $\|u\|_{\infty}=\max\{u,v\}$. Since  $\tau>\ceil{\tau_c}\geq \tau_c$,  $a<b\beta$ by \eqref{arel} and therefore $\alpha<1$. By the contraction mapping theorem (Theorem 2 in \cite{LIZ2002}), $\tilde{z}_0$ is globally asymptotically stable.

If $\tau=\tau_c$, then $a= b\beta$. By the proof in Theorem \ref{Tmlocal} {\it a) ii)}, the solutions remain bounded. We point out that in the same proof, $H$ was shown to be always increasing in the second variable. However,  in this case, $H$ is not necessarily increasing in the first variable. Nevertheless, due to the boundedness  and positivity of $z_t$, there exists a finite  $\bar{z}=\limsup z_t\geq 0$. To show that $\tilde{z}_0=\bar{z}$ and therefore $\tilde{z}_0$ is attractive, we proceed using proof by contradiction. Suppose $\bar{z}>0$. Then 
\begin{align*}
\bar{z} &= \limsup z_{t+1} = \limsup H(z_t,z_{t-\tau+1})\leq \limsup H(z_t, \bar{z}) \\
&= \limsup \left\{\frac{z_t}{1+b+cz_t} + \frac{b^2 \beta \bar{z}}{(1+b+cz_t)[b \beta  + (\beta-1)c\bar{z}]}\right\}\\ 
&\leq \limsup \left\{\frac{z_t}{1+b+cz_t}\right\} + \frac{b^2 \beta \bar{z}}{(1+b)[b \beta  + (\beta-1)c\bar{z}]}\\
&\leq  \frac{\bar{z}}{1+b} + \frac{b^2 \beta \bar{z}}{(1+b)[b \beta  + (\beta-1)c\bar{z}]}
=\frac{b \beta (1+b) + (\beta-1)c\bar{z} }{(1+b)[b \beta  + (\beta-1)c\bar{z}]}\bar{z}< \bar{z},
\end{align*}
contradicting the assumption that $\bar{z}>0$. Therefore, $0\leq \liminf z_t \leq \limsup z_t =0$, and hence $\lim_{t\to \infty}z_t=0$.
\end{proof}

\begin{proof}[\textbf{Proof of Proposition \ref{alowbnd}}]
By \eqref{FP},
\begin{align*}
c\tz \leq 1& \quad \iff \quad -b(2\beta-1)+\sqrt{b^2(2\beta-1)^2 -4b(b\beta-a)(\beta-1)}\leq 2(\beta-1)\\
& \quad \iff \quad -b(2\beta-1)+\sqrt{b(b+4a(\beta-1))}\leq 2(\beta-1)\\
&\quad \iff \quad 4ab(\beta-1)+b^2\leq (2(\beta-1)+b(2\beta-1))^2\\
&\quad \iff \quad 4ab(\beta-1)\leq 
4(\beta-1)^2 + 4b(\beta-1)(2\beta-1) + b^2 (4\beta(\beta-1) +1)-b^2\\
&\quad \iff \quad ab\leq 
(\beta-1) + b(2\beta-1) + b^2 \beta \quad \iff \quad ab\leq  (1+b)(\beta-1+b\beta),
\end{align*}
and the result follows. 
\end{proof}

\begin{proof}[\textbf{Proof of Lemma \ref{mono}}]
Define $w_t:=z_t-\tilde{z}_+$, then 
\begin{equation*}
w_{t+1} = H(z_{t}, z_{t-\tau+1})-\tilde{z}_+= 
H(z_{t}, z_{t-\tau+1})-H(\tilde{z}_+,\tilde{z}_+)
\end{equation*}
Let $u:=z_{t}$ and $v:=z_{t-\tau+1}$.
Then
\begin{align}\label{Useful}
w_{t+1}&=\frac{(1+b)(u-\tilde{z}_+)}{m(u)m(\tilde{z}_+)}  + \frac{ab(b \beta  (1+b)(v-\tilde{z}_+) + b \beta c \tilde{z}_+(v-u)+(\beta-1)c^2\tilde{z}_+v(\tilde{z}_+-u))}{m(u)M(v)m(\tilde{z}_+)M(\tilde{z}_+)}\notag\\
&=\left\{\frac{(1+b)}{m(u)m(\tilde{z}_+)}  
-ab\frac{\overbrace{(b \beta c \tilde{z}_+ +(\beta-1)c^2\tilde{z}_+v)}^{=c\tilde{z}_+M(v)}}{m(u)M(v)m(\tilde{z}_+)M(\tilde{z}_+)}\right\}(u-\tilde{z}_+)+
ab\frac{\overbrace{(b \beta  (1+b)+b \beta c \tilde{z}_+)}^{=b\beta m(\tilde{z}_+)}(v-\tilde{z}_+)}{m(u)M(v)m(\tilde{z}_+)M(\tilde{z}_+)}\notag\\
&=\left\{\frac{1+b}{m(u)m(\tilde{z}_+)}  
-\frac{ abc \tilde{z}_+}{m(u)m(\tilde{z}_+)M(\tilde{z}_+)}\right\}(u-\tilde{z}_+)+\frac{ab^2 \beta }{m(u)M(v)M(\tilde{z}_+)}(v-\tilde{z}_+)\notag\\[2mm]
& \stackrel{\eqref{E_E}}{=}\frac{1-c\tilde{z}_+}{m(u)}  (u-\tilde{z}_+)+
\frac{b \beta (m(\tilde{z}_+)-1) }{m(u)M(v)}(v-\tilde{z}_+)=d_1 w_t + d_2 w_{t-\tau+1},
\end{align}
for $d_1 = \frac{1-c\tz}{m(\tz+w_{t})}$ and $d_2=\frac{b\beta (m(\tz)-1)}{m(\tz+w_t)M(\tz+w_{t-\tau+1})}$. Since $0<c\tilde{z}_+\leq 1$ and $m(\tilde{z}_+)-1=b+c\tilde{z}_+>0$, $d_1,d_2>0$. Hence, if $\mbox{sign}(w_t)=\mbox{sign}(w_{t-\tau+1})$, then $\mbox{sign}(w_{t+1}) = \mbox{sign}(w_{t})$. Consequently, 
$z_{t+1}\geq \tilde{z}_+$ for $z_t,z_{t-\tau+1}\geq \tilde{z}_+$.  Similarly, if $z_t,z_{t-\tau+1}\leq \tilde{z}_+$, then $z_{t+1}\leq \tilde{z}_+$. 
\end{proof}

\begin{proof}[\textbf{Proof of Lemma \ref{Bound}}]
As before, define $w_t:=z_t-\tilde{z}_+$. To show that  \eqref{eq:bound} holds, it suffices to show that 
\begin{equation}\label{eq:bound_w}\min\{ w_0,w_1,\ldots, w_{\tau-1},0\}\leq w_t\leq \max\{w_0,w_1,\ldots, w_{\tau-1}, 0\} \quad \mbox{for all} \quad t\geq \tau.
\end{equation}

First we  consider the lower bound. For any $t\geq\tau$, let $u=z_{t-1}$ and $v =z_{t-\tau}$.
Then by \eqref{Useful} and the fact that $d_1\geq 1+b>0$ and $d_2\geq b \beta>0$, we have
\begin{align*}
w_{t} &=d_1  w_{t-1}+d_2w_{t-\tau}
\geq (d_1+d_2)\min\{w_{t-1},w_{t-\tau},0\}
\geq \left|\frac{1-c\tilde{z}_+}{1+b}  +
\frac{b \beta (m(\tilde{z}_+)-1) }{(1+b)b\beta }\right|\min\{w_{t-1},w_{t-\tau},0\}\\[2mm] 
&= |1|\min\{w_{t-1},w_{t-\tau},0\}
\geq \min\{w_{t-1},w_{t-2}, \ldots,  w_{t-\tau}, 0\}.
\end{align*}
Hence, the lower bound in  \eqref{eq:bound_w}     holds for $t=\tau$. Arguing inductively, it then also  holds for all $t\geq \tau.$
The argument to show the upper bound in  \eqref{eq:bound_w} is similar. Hence, the result follows.
\end{proof}

\begin{proof}[\textbf{Proof of Theorem \ref{TmGlobpos}}]
 As in Lemma~\ref{Bound}, define  $w_t:=z_t-\tilde{z}_+$. Let   $\overline{w}=\limsup w_t$ and $\underline{w}=\liminf w_t$.
Then, by  Lemma \ref{Bound}, both $\overline{w}$ and $\underline{w}$ are finite.
Recalling that $u=z_t$ and $v=z_{t-\tau+1}$ in \eqref{Useful}, we define $\widetilde{H}$ as follows
$$w_{t+1}=H(z_t,z_{t-\tau+1})-H(\tilde{z}_+,\tilde{z}_+)
=\frac{1-c\tilde{z}_+}{m(\tilde{z}_++w_t)}  w_t+\frac{b \beta (m(\tilde{z}_+)-1) }{m(\tilde{z}_++w_t)M(\tilde{z}_++w_{t-\tau+1})} w_{t-\tau+1}:=\widetilde{H}(w_t,w_{t-\tau+1}).$$
Then
$$\frac{\partial \widetilde{H}}{\partial w_{t-\tau+1}}=\frac{b \beta (m(\tilde{z}_+)-1) (b \beta + (\beta-1) c \tilde{z}_+)}{(1 + b + 
   c (\tilde{z}_++w_t)) (b \beta + (\beta-1 ) c (\tilde{z}_++w_{t-\tau+1}))^2}>0.$$
We proceed using proof by contradiction.  Suppose $\overline{w}> 0$. Then
\begin{align*}
\overline{w} &= \limsup w_{t+1} \leq \limsup \widetilde{H}(w_t,\overline{w}) 
    = \limsup \frac{1-c\tilde{z}_+}{1+b+cz_t}w_t + \frac{b \beta  (m(\tilde{z}_+)-1))}{(1+b+cz_t)(b \beta  + (\beta-1)c(\tilde{z}_++\overline{w}))}\overline{w}\\
     &\leq \limsup \frac{1-c\tilde{z}_+}{1+b+cz_t}\overline{w} + \frac{b \beta  (m(\tilde{z}_+)-1))}{(1+b+cz_t)(b \beta  + (\beta-1)c(\tilde{z}_++\overline{w}))}\overline{w}\\
        &\leq  \left(\frac{1-c\tilde{z}_+}{1+b}+\frac{b \beta  (m(\tilde{z}_+-1))}{(1+b)(b \beta  + (\beta-1)c(\tilde{z}_++\overline{w}))}\right)\overline{w}
= \left(\frac{b \beta (1+b) + \overbrace{(1-c\tilde{z}_+)}^{\in (0,1]}(\beta-1)c(\tilde{z}_++\overline{w})}{(1+b)(b \beta  + (\beta-1)c(\tilde{z}_++\overline{w}))}\right)\overline{w}\\
&<\overline{w},
\end{align*}
contradicting the assumption that $\overline{w}>0$.
Hence, $\overline{w}\leq 0$ and therefore  $\underline{w}\leq 0$.

Next we show that  $\underline{w}=0$.   
Since $z_t\geq 0$, $\underline{w}\geq -\tilde{z}_+$.
Again we proceed using proof by contradiction.  Suppose $\underline{w}=-\tilde{z}_+$. Then there must exist a subsequence $w_{t_j}$ converging to $-\tilde{z}_+$. By Lemma \ref{Lpos} for $\vec{z}_0\not \equiv \vec{0}$, we can assume, without loss of generality, that \ $z_i>0$, i.e., $w_i>-\tilde{z}_i$ for $i=0,1,\ldots, \tau-1$.   By Lemma \ref{Bound}, $w_t\geq \min\{w_0,w_1,\ldots, w_{\tau-1},0\}>-\tilde{z}_+$, violating the assumption that the  subsequence decreases to $-\tilde{z}_+$. Suppose therefore $-\tilde{z}_+<\underline{w}<0$. Then 
\begin{align*}
    \underline{w} &=\liminf w_{t+1} \geq \liminf \widetilde{H}(w_t,\underline{w}) =\liminf \frac{1-c\tilde{z}_+}{1+b+cz_t} w_t + \frac{b \beta  (b+c\tilde{z}_+)}{(1+b+cz_t)(b \beta  + (\beta-1)c(\tilde{z}_++\underline{w}))}\underline{w}\\
    &\geq \liminf \frac{1-c\tilde{z}_+}{1+b+cz_t} \underline{w} + \frac{b \beta  (b+c\tilde{z}_+)}{(1+b+cz_t)(b \beta + (\beta-1)c(\tilde{z}_++\underline{w}))}\underline{w}\\
    & \stackrel{\underline{w}<0}{\geq} \left(\frac{1-c\tilde{z}_+}{1+b}  + \frac{b\beta  (b+c\tilde{z}_+)}{(1+b)(b \beta + (\beta-1)c(\tilde{z}_++\underline{w}))}\right)\underline{w}
    =\frac{b \beta (1+b) +(1-c\tilde{z}_+) (\beta-1)c(\tilde{z}_++\underline{w})
   }{(1+b)(b \beta + (\beta-1)c(\tilde{z}_++\underline{w}))}\underline{w}> \underline{w},
\end{align*}
because $c\tilde{z}_+\in (0,1]$. This   violates the assumption that $\underline{w}<0$. Therefore,   $\underline{w}=0=\overline{w}$, completing the proof.
\end{proof}

 For the reader's convenience, we state the following theorem from \cite{Ladas_2004} that we will use in our proof  of Theorem \ref{Thm_glob_pos2}, where we prove global stability of the positive equilibrium.\\[2mm] 

\noindent \textbf{Theorem~1.15 in \cite{Ladas_2004}}\label{Ladas}
{\it Let $g:[a,b]^{k+1} \to [a,b]$ be a continuous function, where $k$ is a positive integer and $[a,b]$ is an interval of real numbers. Consider 
$$x_{n+1} = g(x_n,x_{n-1},\ldots, x_{n-k}),\quad \quad n=0,1,\ldots$$
Assume that $g$ satisfies the following conditions: 
\begin{enumerate}
    \item For each integer $i$ with $1\leq i\leq k+1$, the function $g(z_1,z_2,\ldots, z_{k+1})$ is weakly monotonic in $z_i$ for fixed $z_j$, $j\neq i$.
    \item If $(r,R)$ is a solution of the system
    $$r=g(r_1,r_2,\ldots, r_{k+1}), \quad \quad R=g(R_1,R_2,\ldots, R_{k+1})$$
    then $r=R$, where for each $i=1,2,\ldots, k+1$, we set
    $$R_i = \begin{cases} R & \mbox{ if g is non-decreasing in }z_i\\
    r & \mbox{ if g is non-increasing in }z_i\end{cases}$$
    and 
    $$r_i = \begin{cases} r & \mbox{ if g is non-decreasing in }z_i\\
    R & \mbox{ if g is non-increasing in }z_i\end{cases}$$
\end{enumerate}
Then there exists exactly one equilibrium $\bar{x}$ and every solution converges to $\bar{x}$.
}

\begin{proof}[\textbf{Proof of Proposition \ref{Lowerbnd}}]
 Differentiating $H(z_t,z_{t-\tau+1})$ with respect to $z_t$ is given by  \eqref{eq:partialH1}. Simplifying yields
    $$\frac{\partial H}{\partial z_t}(z_t,z_{t-\tau+1})=\frac{b \beta(1+b)  + c( (\beta-1)(1+b)-ab)z_{t-\tau+1}}{(1+b+c z_t)^2(b \beta + (\beta - 1)c z_{t-\tau+1})}.$$
  Since the denominator is positive,
$$\frac{\partial H}{\partial z_t}(z_t,z_{t-\tau+1}) >0 \quad \iff 
\quad P(z_{t-\tau+1})=a_0+a_1z_{t-\tau+1}>0$$
where
\begin{equation}\label{a0a1}
a_0=b\beta(1+b),\quad \quad a_1 = c((\beta-1)(1+b)-ab) = c((\beta-1)(1+b)-(b+c\tz)(\beta b + (\beta-1)c\tz)),
\end{equation}
and we replaced   $a b$ using \eqref{E_E}.
Clearly, $a_0>0$ and since we are assuming that $c\tz>1$, it follows that $a_1<0$,  since
\begin{eqnarray}
a_1&=&c((\beta-1)(1+b)-(b+c\tz)(b \beta  + (\beta-1)c\tz)) \nonumber\\
&< &
c((\beta-1)(1+b)-(b+1)(\beta (b+1) -1)) \nonumber\\
&=& -cb\beta(1+b)<0. \label{eq:a1neg}
\end{eqnarray}
Therefore, 
\begin{equation}\label{Hinc}
\frac{\partial H}{\partial z_t} \begin{cases}
>0 & z_{t-\tau+1}<\chi:=\frac{a_0}{(-a_1)},\\
<0 & z_{t-\tau+1}>\chi,\\
=0 & z_{t-\tau+1} = \chi.
\end{cases}
\end{equation}
Since $c\tz>1$, 
\begin{equation*}
\beta b < c\tz b \beta < c\tz (b\beta+(\beta-1)(c\tz-1))
\end{equation*}
and therefore,
\begin{align*}
\beta b (1+b)&<c\tz (1+b)(b \beta + (\beta-1)(c\tz-1))< c\tz ( (1+b)(b \beta + (\beta-1)c\tz) - (\beta-1)(1+b))\\
&<  c\tz ( (b+c\tz)(b \beta + (\beta-1)c\tz) - (\beta-1)(1+b))
\end{align*}
Dividing by the right-hand side yields
\begin{equation}\label{importantchi}
\chi=\frac{\beta b (1+b)}{c((b+c\tz)(b \beta + (\beta-1)c\tz) - (\beta-1)(1+b))}<\tz.
\end{equation}
Further, since $\chi = \frac{a_0}{-a_1}$,  by \eqref{a0a1}, 
\begin{align*}
c\chi<1  &  \quad \iff \quad c\frac{b\beta(1+b)}{c\left[ (b+\tz)(\beta b + (\beta-1)c\tz) - (\beta-1)(1+b)\right]}<1 \\[2mm]
& \quad \iff \quad b\beta(1+b)< (b+\tz)(\beta b + (\beta-1)c\tz) - (\beta-1)(1+b)\\[2mm]
& \quad \iff \quad b\beta(1+b)+ (\beta-1)(1+b)< (b+\tz)(\beta b + (\beta-1)c\tz). 
\end{align*}
The last inequality holds, since  for $c\tz>1$  
$$ (b+1)(b\beta+(\beta-1))< (b+c\tz)(b\beta+(\beta-1))< (b+c\tz)(\beta b + (\beta-1)c\tz).$$
 Combining this with  \eqref{importantchi}, we have
\begin{equation}\label{importantc}
    c\chi<1  \qquad \mbox{ and } \qquad \chi<\tz.
\end{equation}

We now claim that 
\begin{equation}\label{Aim1}
    z_{t+1}=H(z_t,z_{t-\tau+1})> \begin{cases} \chi & \mbox{ if } z_{t-\tau+1}\geq \chi,\\
    \min\{z_t,z_{t-\tau+1}\} & \mbox{ if } z_{t-\tau+1}<\chi.
    \end{cases}
\end{equation}

Firstly, we show that if $z_{t-\tau+1}\geq \chi$, then  $z_{t+1}>\chi$. Since
\begin{equation*}
\lim_{Z\to \infty}H(Z,\chi) = \lim_{Z \to \infty} \frac{ZM(\chi)+(m(\tz)-1)M(\tz)\chi}{(1+b+cZ)M(\chi)}=\frac{1}{c}> \chi,
\end{equation*}
and 
 by \eqref{Hinc},  $H$ is non-increasing in the first variable,
 so for any $Z>z_t$, we have 
\begin{equation*}
\chi<\frac{1}{c}\leq H(Z,\chi)\leq H(z_t,\chi)\leq H(z_t,z_{t-\tau+1}) ,
\end{equation*}
where the last inequality holds, because $H$ is increasing in the second variable.

Secondly, if $z_{t-\tau+1}<\chi$, we show that \eqref{Aim1} holds by showing that
\begin{equation*}
    z_{t+1}>z_m, \quad \mbox{where} \quad z_m = \min\{z_t,z_{t-\tau+1}\}.
\end{equation*}
This inequality is satisfied, since for $z_{t-\tau+1}<\chi<\tz$,  $H$ is strictly increasing in both variables, and therefore
\begin{eqnarray*}
    z_{t+1}&=&H(z_t,z_{t-\tau+1})\geq H(z_m,z_m)= \frac{M(z_m)+(m(\tz)-1)M(\tz)}{m(z_m)M(z_m)}z_m \notag\\
    &>&\frac{M(z_m)+(m(z_m)-1)M(z_m)}{m(z_m)M(z_m)}z_m=z_m,
\end{eqnarray*}
which results in \eqref{Aim1}. 

We now define
\begin{equation}\label{Defzmi}
z_{m_i} := \min\Big\{z_t \, | \, i \tau\leq  t\leq  (i+1)\tau-1   \Big\}, \quad i=0,1,2,\dots,
\end{equation}
and use  \eqref{importantc} and \eqref{Aim1}  to prove that $z_t>\chi$  for all sufficiently large $t$. 
We proceed using proof by contradiction. 
Suppose this is not true. Then for every fixed $t\geq 0$, there exists $\hat{T}>t$ such that $z_{\hat{T}}\leq \chi$. By \eqref{Aim1}, this implies that $z_{m_i}<\chi$ for all $i$. 
Since the sequence $\{z_{m_i}\}$ is also increasing, there exists $z^*$ such that
\begin{equation}\label{Contra}
    z^*=\lim \limits_{i \to \infty} z_{m_i}\leq \chi.
\end{equation}
Then, for each  $\epsilon>0$ there exists $j(\epsilon)$ such that $z^*-\epsilon< z_{m_{j(\epsilon)}}\leq z^*$. By \eqref{Defzmi}, this also implies that
\begin{equation}\label{Contra3}
z^*-\epsilon < z_t \qquad \mbox{ for all } \quad t \geq j(\epsilon) \tau
\end{equation}
and 
\begin{equation}\label{Contra2}
S_i=\{z_t \, | \, z_t\leq z^*, \quad t\geq i \tau\} \neq \emptyset, \quad \quad \mbox{ for all } i \geq j(\epsilon).
\end{equation}
By \eqref{Aim1}, if $z_s,z_{s+\tau-1} \notin S_i$, then $z_{s+\tau}\notin S_i$. 
Let $z_s \in S_i$, for some fixed but arbitrary $i\geq j(\epsilon)$. Then $z_s\leq z^*$, and by \eqref{Contra3}, $z^*-\epsilon<z_s,z_{s+\tau-1}$. Since $z_s\leq z^* \leq \chi$, $H$ is non-decreasing in both variables and we obtain
$$z_{s+\tau}=H(z_{s+\tau-1}, z_{s}) \geq  H(z^*-\epsilon,z^*-\epsilon).$$

We obtain a contradiction to \eqref{Contra2} by showing that   there exists $\epsilon \in (0,z^*)$ such that $H(z^*-\epsilon,z^*-\epsilon)>z^*$, since then $z_{s+\tau}>z^*$, and this implies that $S_i = \emptyset$ for $i>j(\epsilon)+1$. 
To show the existence of such an $\epsilon \in (0,\tz)$,   note that $H(z^*-\epsilon,z^*-\epsilon)-z^*$ is of the form
$$  H(z^*-\epsilon,z^*-\epsilon)-z^*=\frac{\alpha_2 \epsilon^2  + \alpha_1\epsilon + \alpha_0}{m(z^*-\epsilon)M(z^*-\epsilon)}.$$
Since the denominator is positive and 
$$\alpha_0 = c z^* (\tz - z^*)  [b (2 \beta-1) + c(\beta-1)  (\tz + z^*)]>0,$$
there exists $\delta>0$ such that for $\epsilon \in (0,\delta)$, $\alpha_2\epsilon^2 + \alpha_1 \epsilon + \alpha_0>0$. Hence, there exists $\epsilon\in (0,z^*)$ such that $H(z^*-\epsilon, z^*-\epsilon)>z^*$.
Therefore, we have obtained a contradiction and so there exists $T$ such that $z_t\geq \chi$ for all $t\geq T$.
\end{proof}

\begin{proof}[\textbf{Proof of Proposition \ref{invariant}}]
By Proposition \eqref{Lowerbnd}, there exists $T$ such that $z_t \geq \chi$ for all $t \geq T$. Without loss of generality, let $T=0$. We prove that there  exists $U>\tz$ such that $z_{t+1}\in [\chi,U]$ for $z_t,z_{t-\tau+1} \in [\chi,U]$. Since $H$ is increasing in the second variable, and by Proposition \eqref{Lowerbnd}, $z_t\geq \chi$ for all $t\geq 0$, it follows using by \eqref{Hinc} that $H$ is decreasing in the first variable.  
Therefore, if such a $U$ exists, then
$$z_{t+1} = H(z_t,z_{t-\tau+1})\leq H(\chi,U), \quad \quad z_t,z_{t-\tau+1} \in [\chi,U].$$
Hence, to prove the existence of such a $U$, it suffices to show that $H(\chi,U)\leq U$ for some $U>\tz$.
\begin{equation}\label{Qeq}
  H(\chi,U)-U =\frac{Q(U)}{(1 + b) c m(\tz)M(U) ( b \beta + (\beta-1) (c \tz-1)) }
\end{equation}
where $Q(U)$ is a second-order polynomial of the form $q_2 U^2 + q_1 U + q_0$, with
$$q_2 = -c^2 (1 + b) (\beta-1) m(\tz) ( b \beta+(\beta-1) (c \tz-1))<0.$$
Therefore, there exists $U^*>\tz>\chi$ such that $Q(U^*)\leq 0$ and therefore, since the denominator in \eqref{Qeq} is positive, $H(\chi,U)\leq U$, for all $U\geq U^*$, completing the proof. 
\end{proof}

\begin{proof}[\textbf{Proof of Theorem \ref{Thm_glob_pos2}}]
Let $\vec{z}_0 \not \equiv \vec{0}$. Then, by Proposition \ref{Lowerbnd}, there exists $T$ such that $z_t\geq \chi$ for all $t \geq T$. By Proposition \ref{invariant}, there exists $U>\tz$ such that $z_t \in [\chi,U]$ for $t \geq T$. Without loss of generality, we therefore assume $z_s \in [\chi,U]$ for $s =0, 1, \ldots, \tau-1$, and $H:[\chi,U]\times [\chi,U] \to [\chi,U]$. In that case, $H$ is decreasing in the first variable and strictly increasing in the second variable, hence satisfying {\it 1)} in \cite[Theorem 1.15]{Ladas_2004}.

Next  we show that {\it 2)} in \cite[Theorem 1.15]{Ladas_2004} also holds to obtain the result.
Consider now $r,R \in [\chi,U]$ such that
\begin{align}
r&=H(R,r)=\frac{R}{m(R)} + \frac{abr}{m(R)M(r)}\label{Hope1}\\
R&=H(r,R)=\frac{r}{m(r)} + \frac{abR}{m(r)M(R)}.\label{Hope2}
\end{align}
In what follows, we show that $r=R=\tz$ is the only solution to \eqref{Hope1}--\eqref{Hope2} in $[\chi,U]$. To find all possible solutions to \eqref{Hope1}--\eqref{Hope2}, we multiply \eqref{Hope1} by its denominator and obtain
\begin{equation}\label{Rone2}
rM(r)m(R) = RM(r)+abr.
\end{equation}
Solving for $M(r)$, we obtain
\begin{equation}\label{Rone}
M(r)  = \frac{abr}{r(1+b)+R(cr-1)}.
\end{equation}
If $cr=1$, then \eqref{Rone} reduces to 
$$ b \beta  + (\beta-1) = \frac{ab}{1+b},$$ 
which violates \eqref{acause}, since $c\tz>1$. 
Therefore, $cr\neq 1$. In this case, we solve \eqref{Rone2} for $R$ and obtain
\begin{equation}\label{M1}
    R =r \frac{ ab-M(r)(1+b)}{(cr-1)M(r)}.
\end{equation}
If $cr< 1$, or equivalently $r< 1/c$, then, by \eqref{acause}, 
$$(1+b)M(r)\leq (1+b)M(1/c) =(1+b)(b\beta  +(\beta-1))<ab.$$
This results in a negative value on the right-hand side of \eqref{M1}, which violates the condition that $R \in [\chi,U]$. The only possibility that remains is that 
\begin{equation}\label{eq:cr}
    cr>1.
\end{equation}

Next we find the solutions of \eqref{Hope1}--\eqref{Hope2}. Rearranging terms in \eqref{Hope2} and solving for $R$ yields
\begin{align*}
Rm(r)M(R)=rM(R) + abR
 \quad &\iff \quad Rb \beta m(r) +c(\beta-1) m(r)R^2=rb\beta +(\beta-1)crR + abR\\
&\iff \quad R^2\{ c(\beta-1)m(r)\} + R\left\{b \beta m(r)-(\beta-1)cr - ab \right\}-rb\beta =0.
\end{align*}
Since $c(\beta-1)m(r)>0$ and $r\beta b>0$, there exists exactly one positive root given by
\begin{equation}\label{Mplus}
R_+ =\frac{(\beta-1)cr + ab-b\beta  m(r)+ \sqrt{\left\{b\beta  m(r)-(\beta-1)cr - ab \right\}^2 + 4 rb\beta  c(\beta-1)m(r)}}{2c(\beta-1)m(r)}.
\end{equation}
Since  the values of $R$ in \eqref{M1} and \eqref{Mplus} must be equal $R-R_+=0$, and hence
\begin{align*}
   0 =R-R_+=\frac{P(r)}{2c (\beta-1)  (cr-1) m(r) M(r)}  \,\iff \,  P(r)=0,
\end{align*}
where $P(r)=0$ if and only if  
\begin{multline*}
r \left\{ ab-M(r)(1+b)\right\}2c(\beta-1)m(r)-(cr-1)M(r)\left\{(\beta-1)cr + ab-b\beta m(r)\right\}\\
= (cr-1)M(r)\sqrt{\left\{b \beta  m(r)-(\beta-1)cr - ab \right\}^2 + 4  rb\beta  c(\beta-1)m(r)}.
\end{multline*}
If there exists $r$ such that this equality is satisfied, then $r$ also solves 
\begin{multline*}
\left(r \left\{ ab-M(r)(1+b)\right\}2c(\beta-1)m(r)-(cr-1)M(r)\left\{(\beta-1)cr + ab-b \beta  m(r)\right\}\right)^2\\
- (cr-1)^2M^2(r)\left( \left\{b \beta  m(r)-(\beta-1)cr - ab \right\}^2 + 4 r\beta b c(\beta-1)m(r)\right)=0.
\end{multline*}
Since $m(r)$ and $M(r)$ are linear functions in $r$, the left-hand side can be expressed as a sixth order polynomial in $r$, namely  
\begin{align*}
   & \hat{P}(r)=\sum_{i=0}^6 \alpha_i r^i \quad \quad \quad \mbox{ with }\\
\alpha_0&=0\\
\alpha_1&=4c b^3\beta (1 + b) (\beta-1)  (a^2 - 2 a \beta (1 + b) +  b \beta^2(2 + b))\\
\alpha_2&= -4 b^2c^2 (\beta-1) (a^2 (2 + b^2 - 3 b (\beta-1) - 2 \beta) -   a  \beta(1 + b) (4 + 2 b^2 + b (7 - 8 \beta) - 4 \beta) \\
& \hspace{20mm}+ b \beta^2 (2 + b) (3 + b^2 + b (4 - 5 \beta) - 4 \beta) )\\ 
\alpha_3&=-4 bc^3  (\beta-1) (a^2 b (-2 \beta+2+b) + 
   b \beta (2 + b) (3 + 2 b^2 - 5 b (\beta-1) - 3 \beta)  (2 \beta-1) \\
   &\hspace{20mm}+  a (b^3 (2 - 5 \beta) + 2 (\beta-1)^2 + b^2 (7 - 21 \beta + 12 \beta^2) + 
      b(\beta-1) (-7+ 12 \beta)))\\
    \alpha_4&=  4 b c^4 (\beta-1) (a (-4 (\beta-1)^2 - 8 b (\beta-1)^2 + b^2 (-3 + 4 \beta)) - (2 + b) (-(\beta-1)^2 (-1 + 4 \beta) \\
    &\hspace{20mm}+ b^2 (1 - 6 \beta + 6 \beta^2) + 
      b (2 - 13 \beta + 21 \beta^2 - 10 \beta^3)))\\
      \alpha_5&=4c^5 (\beta-1)^2 (b^3 2(1 - 2 \beta) - 
   b (7 + 2 a - 11 \beta) (\beta-1) + 2 (\beta-1)^2 + 
   b^2 (7 + a - 16 \beta + 5 \beta^2)) \\
   \alpha_6&=4 (2 + b) (\beta-(1 + b)) (\beta-1)^3 c^6.
\end{align*}
The first three roots are easily found as equilibria of \eqref{eq10}. The other roots are obtained using the symbolic computing environment in  Maple \cite{maple} and  can be checked analytically:
\begin{equation}\label{sixroots}
    r_1 = 0, \quad  r_2 = \tz, \quad r_3=\tilde{z}_{-},  \quad  r_4 = \frac{-(1+b)}{c} \quad   r_5 =\frac{\gamma_1+\sqrt{\gamma_2}}{\gamma_3}, \quad  r_6 = \frac{\gamma_1-\sqrt{\gamma_2}}{\gamma_3}\\
    \end{equation}
    where
    \begin{align*}
\gamma_3&=-2 c^2 (2 + b) (\beta-(1+b)) (\beta-1) <0,\\[2mm]
\gamma_1&= c\left\{-(c\tz)^2 (\beta-1)(2\beta-2-b)+(c\tz) b\left[ (\beta-1)(b+2(1-2\beta))+b\beta\right] + 2b\beta [2\beta-2-b]\right\},\\[2mm]
         \gamma_2&=\gamma_1^2+4 b\beta c^2 (2 + b) (\beta-1-b) (\beta-1)  (  c\tz [ (\beta-1) c \tz-b] + 2 b\beta  (c \tz-1)).
\end{align*}
Clearly, $r_1, r_3, r_4 \notin [\chi,U]$, since they are negative. We next show that  $r_5$ and $r_6$ are also both not feasible. Interpreting $\gamma_1$ as a function of $c\tz$, $$\gamma_1=\gamma_1(c\tz)=\hat{a}_2(c\tz)^2+\hat{a}_1(c\tz) + \hat{a}_0, \qquad  \hat{a}_0=2b\beta c(2\beta-2-b)>0.$$ 
    Since $\hat{a}_2= -c(\beta-1)(2\beta-2-b)<0$,  there exists exactly one positive root and one negative root of the equation $\gamma_1=0$. This root lies in the interval $(0,1)$, because  
\begin{equation*}
\gamma_1(0)=\hat{a}_0>0 \qquad \mbox{and} \qquad    \gamma_1(1)= -c(\beta-1-b)( 2\beta-2-b) <0,
\end{equation*}
and so $\gamma_1<0$ for all $c\tz>1$. Also, since $\gamma_3<0$, the roots $r_5$ and $r_6$ can be expressed as 
\begin{equation}\label{eq:bad_roots}
r_5 = \frac{|\gamma_1|-\sqrt{\gamma_2}}{|\gamma_3|} \quad \quad \mbox{and}\quad \quad r_6=\frac{|\gamma_1|+\sqrt{\gamma_2}}{|\gamma_3|}.
\end{equation}

It follows  that $r_5$ is negative and hence  not feasible, since
$$\gamma_2=\gamma_1^2 + 4b(2+b)(\beta-1-b)(\beta-1)\beta c^2 (c \tz[(\beta-1)c\tz-b]+2b\beta (c\tz-1))> \gamma_1^2,$$
and $\beta>1+b$, $c\tz>1$ and $(\beta-1)c\tz-b>(\beta-1)-b>0$. 

Next we show  that although $r_6$ and the corresponding $R_6$ value solve \eqref{Hope1}--\eqref{Hope2}, if  $r_6\in [\chi,U]$, then $R_6 <0$ and hence not feasible. We proceed using proof by contradiction.
If $r_6 \in [\chi,U]$, then by \eqref{eq:cr} $cr_6>1$. By \eqref{M1} and \eqref{eq:bad_roots}, $R_6>0$ if
and only if
\begin{align}
&ab>(1+b)M(r_6) \quad \iff \quad
(m(\tz)-1)M(\tz)>(1+b)M(r_6)\notag \\
&\iff \quad  
(b+c\tz)(\beta b + (\beta-1)c\tz)>(1+b)\left[ \beta b + (\beta-1)c\frac{|\gamma_1|+\sqrt{\gamma_2}}{|\gamma_3|}\right]\notag \\
& \iff \quad 
\frac{|\gamma_3|}{c}\left[(b+c\tz)(\beta b + (\beta-1)c\tz)-(1+b)\beta b\right] - (1+b)(\beta-1)|\gamma_1|>(1+b)(\beta-1)\sqrt{\gamma_2}.\label{InR5}
\end{align}
Since  the right hand side of \eqref{InR5} is positive,  both sides must be positive and so we can square both sides to obtain:
\begin{align}
&\frac{|\gamma_3|^2}{c^2}\left[(b+c\tz)(\beta b + (\beta-1)c\tz)-(1+b)\beta b\right]^2+ (1+b)^2(\beta-1)^2|\gamma_1|^2\notag \\ 
& \hspace{15mm} -2(1+b)(\beta-1)|\gamma_1|\frac{|\gamma_3|}{c}\left[(b+c\tz)(\beta b + (\beta-1)c\tz)-(1+b)\beta b\right]\notag\\
& \hspace{3cm}>(1+b)^2(\beta-1)^2\gamma_2\notag\\
&\iff \quad \quad  Q(c\tz):= b_0 + b_1(c\tz) + b_2 (c\tz)^2 + b_3(c\tz)^3 + b_4(c\tz)^4>0 \label{Qpos}
\end{align}
where 
\begin{align*}
    b_0&= 4 b^3\beta^2 c^2(2 + b) (\beta-1)^2  \left[(\beta-1)^2(1 + 
   2 b) - b^2 (2 \beta-1)\right] \\
   b_1&= 4 b^2 \beta c^2(2 + b) (\beta-1)^2    (2 \beta-1) [(\beta-1)^2 (1+ 
   2 b) + b^3 \beta - b^2 (-1 + \beta + \beta^2)] \\
   b_2&=  4b\beta c^2(2+b)(\beta-1)^2  [ (\beta-1)^3 (1+ 2 b) + 
   b^2 \beta (-3 + 8 \beta - 5 \beta^2) + 
   b^3 (1 + 5 \beta(\beta-1))]\\
   b_3&=-8 b^2\beta c^2 (2+b)(\beta-1)^3  (\beta-1 - b)   (2 \beta-1) \\
   b_4&= -4b\beta c^2(2+b)(\beta-1)^4  (\beta-1 - b).
\end{align*}
Interpreting the left-hand side of \eqref{Qpos} as a function of $c\tz$, it is a polynomial of order four. Its four roots, obtained by Maple  and easily verifiable, are given by
$$(c\tz)_1 =1,\quad (c\tz)_2=-b, \quad  (c\tz)_3=\frac{-b\beta}{\beta-1},\quad (c\tz)_4= \frac{-[\beta(2b+1)-(1+b)]}{\beta-1}.$$
The largest positive root is $(c\tz)_1=1$. Since $b_4<0$, $Q(c\tz)<0$ for all $c\tz>1$. This violates \eqref{Qpos}, since $c\tz>1$.

Finally, $r_2=\tz$, a fixed point of the system, and hence 
$R_2=\tz$. Therefore, $r_2=R_2 \in [\chi,U]$ and is  
the only feasible solution of \eqref{Hope1}--\eqref{Hope2}. 
By \cite[Theorem 1.15]{Ladas_2004}, $\tz$ is globally asymptotically stable.
\end{proof}

\section*{Acknowledgement}
The research of Gail S. K. Wolkowicz was partially supported by a Natural Sciences and Engineering Research Council of Canada (NSERC) Discovery grant with accelerator supplement.

\bibliographystyle{abbrv}

\bibliography{Lib}

\end{document}